\documentclass[aps,reprint,citeautoscript,superscriptaddress,showpacs]{revtex4-1}

\usepackage{graphicx,epsfig,float}
\usepackage{amssymb}
\usepackage{amsmath}
\usepackage[varg]{txfonts}% times in math

\allowdisplaybreaks
\renewcommand{\vec}[1]{\mathbf{#1}}      %vector --- boldsymbol

\begin{document}
\title{Universal description of potential energy surface of interlayer interaction in two-dimensional materials by first spatial Fourier harmonics}

\author{Alexander V. Lebedev}
\email{allexandrleb@gmail.com}
\affiliation{Kintech Lab Ltd., 3rd Khoroshevskaya Street 12, Moscow 123298, Russia}
\author{Irina V. Lebedeva}
\email{liv\_ira@hotmail.com}
\affiliation{CIC nanoGUNE BRTA, San Sebasti\'an 20018, Spain}
\author{Andrey M. Popov}
\email{popov-isan@mail.ru}
\affiliation{Institute for Spectroscopy of Russian Academy of Sciences, Troitsk, Moscow 108840, Russia}
\author{Andrey A. Knizhnik}
\email{andrey.knizhnik@gmail.com}
\affiliation{Kintech Lab Ltd., 3rd Khoroshevskaya Street 12, Moscow 123298, Russia}
\author{Nikolai A. Poklonski}
\email{poklonski@bsu.by}
\affiliation{Belarusian State University, Nezavisimosti Ave.~4, Minsk 220030, Belarus}
\author{Sergey A. Vyrko}
\affiliation{Belarusian State University, Nezavisimosti Ave.~4, Minsk 220030, Belarus}

\begin{abstract}
We propose a hypothesis that the potential energy surface (PES) of interlayer interaction in diverse 2D materials can be universally described by the first spatial Fourier harmonics. This statement (checked previously for the interactions between graphene and hexagonal boron nitride layers in different combinations) is verified in the present paper for the case of hydrofluorinated graphene (HFG) bilayer with hydrogen bonding between fluorine and hydrogen at the interlayer interface. The PES for HFG bilayer is obtained through density functional theory calculations with van der Waals corrections. An analytical expression based on the first Fourier harmonics describing the PES which corresponds to the symmetry of HFG layers is derived. It is found that the calculated PES can be described by the first Fourier harmonics with the accuracy of 3\% relative to the PES corrugation. The shear mode frequency, shear modulus and barrier for relative rotation of the layers to incommensurate states of HFG bilayer are estimated. Additionally it is shown that HFG bilayer is stable relative to the formation of HF molecules as a result of chemical reactions between the layers.
\end{abstract}
\maketitle

\section{Introduction}

Since the discovery of graphene \cite{Novoselov2004} this material has attracted considerable attention due to its unique physical properties. The interaction between graphene layers is responsible for the tunable band gap \cite{Zhang2009} and  such phenomena as superconductivity in twisted graphene bilayers \cite{Cao2018}, commensurate-incommensurate phase transition \cite{Popov2011} manifested through formation of a network of domain walls \cite{Lebedeva2019} with topologically protected helical states \cite{Vaezi2013,Zhang2013} in bilayer graphene, self-retraction of graphene layers \cite{Zheng2008,Popov2011a} and so on. Graphene applications based on interlayer interaction such as nanoelectromechanical systems (NEMS) composed of graphene layers which slide with respect to each other have been proposed \cite{Poklonski13JCTN141, Kang2015, Kang2016}.
In addition to graphene, a wide family of other 2D materials has been recently synthesized including hexagonal boron nitride (h-BN, see Ref.~\cite{Auwarter2019} for a review), graphane \cite{Elias2009}, various transition metal dichalcogenides (see Ref.~\cite{Shi2015} for a review), phosphorene \cite{Churchill2014}, borophene \cite{Zhang2015}, germanene \cite{Yuhara2018}, {\it etc}. Heterostructures consisting of layers of different 2D materials should be also mentioned (see, e.g., Ref.~\cite{Siahlo18PRM036001} on graphene/h-BN nanoscrolls and Ref.~\cite{Geim2013} for a review). An important characteristic of interlayer interaction is the potential energy surface (PES) that is the interlayer interaction energy as a function of the coordinates describing the relative displacement of the layers.
Particularly the PES determines the commensurate-incommensurate phase transition in 2D bilayers, self-retraction of 2D layers and operation of NEMS based on relative motion of such layers.

The first-principles calculations for graphene \cite{Ershova2010,Lebedeva2011,Popov2012,Lebedeva2012,Reguzzoni2012} and h-BN \cite{Lebedev2016} bilayers and graphene/h-BN heterostructure \cite{Jung2015,Kumar2015,Lebedev2017} have shown that the PESs of interlayer interaction in such systems can be described using the first spatial Fourier harmonics.
It is interesting to note that while the amplitude of corrugations of the PES for graphene bilayer computed using the simple Lennard-Jones potential is an order of magnitude less than for the surface that follows from the density functional theory (DFT) calculations, it is described excellently by the same expression \cite{Lebedeva2011}.
The approximation by the first Fourier harmonics also works well for PESs of interwall interaction of infinite and commensurate carbon nanotube walls \cite{Vucovic2003,Belikov2004, Bichoutskaia2005, Bichoutskaia2009, Popov2009, Popov2012a} and nanotube walls with edges \cite{Popov2013} and defects \cite{Belikov2004}, both obtained from first principles \cite{Bichoutskaia2005, Bichoutskaia2009, Popov2009, Popov2012a,Popov2013} and using classical potentials \cite{Vucovic2003,Belikov2004}. Based on the results listed above for PESs of interlayer interaction between 2D layers and nanotube walls, we propose here the hypothesis that the possibility of approximation of the PES by the first Fourier harmonics is a universal property for diverse 2D materials. 

This hypothesis leads to the important conclusion that physical properties  determined by different regions of the PES are interrelated, that is the measurement of any physical property determined by the PES gives the information about the whole PES and, therefore, can be used to estimate other properties determined by this PES \cite{Popov2012}. In the case of interaction between graphene layers, the proposed hypothesis is not only based  on the DFT calculations but also confirmed by the following experimental data. The barrier for relative motion of graphene layers derived through measurements of the shear mode frequency (related with the PES near its minimum) and the stacking dislocation width (related with the PES along the path between neighbouring minima through the saddle point) equals 1.5--1.8 meV~\cite{Popov2012} and 2.4~meV~\cite{Alden2013} per carbon atom of one layer (per carbon atom of the upper/adsorbed layer), respectively. The difference between these two estimates of the barrier is less than the scatter of the values that follow from the DFT calculations ranging from 0.5 to 2.4~meV per carbon atom of one layer, see Ref.~\cite{Lebedeva2016a} and references therein. Approximations by Fourier harmonics beyond the first ones have been also considered for graphene \cite{Zhou2015}, h-BN \cite{Zhou2015} and MoS$_2$ \cite{Carr2018} bilayers and graphene/h-BN heterostructure \cite{Zhou2015}. 

The examples where the PES is described using the first Fourier harmonics listed above correspond to the simplest 2D materials, graphene and h-BN. In the present paper, we consider this approximation for the PES of chemically modified graphene layers by the example of interaction between hydrofluorinated graphene (HFG) layers with hydrogen bonding between fluorine and hydrogen atoms at their interface. Bilayer and multilayer systems consisting of chemically modified graphene layers such as graphane, fluorographene and HFG in different combinations have been studied recently because of their interesting electronic properties and possible applications in nanoelectronics \cite{Rohrer2011,Lu2009,Li2012,Tang2014,Kim2015}.
HFG is a so-called janus nanostructure with piezoelectricity within one layer \cite{Ong2013,Kim2014,Cakir2015}. This 2D material has been synthesized \cite{Sofer2015} and the structure and electronic properties of the monolayer have been investigated theoretically in Refs.~\cite{Medeiros2010,Singh2011,Ong2013,Kim2014,Kim2015,Cakir2015,Aggoune2016}. 
The piezoelectric enhancement has been predicted for HFG bilayer \cite{Kim2015}. While the energies of interlayer interaction have been found for some symmetric stackings of hydrofluorinated graphene bilayer \cite{Kim2015,Li2015} and graphane/fluorographene heterostructure \cite{Li2012}, the whole PES has not been yet considered for the systems of chemically modified graphene layers. Based on the PES approximation, we also estimate  properties of HFG bilayer associated with relative sliding of the layers: shear mode frequency, shear modulus and barrier for relative rotation of the layers to incommensurate states. 

The DFT study of the interaction between HFG layers allows us to consider an additional problem. The chemical modification of 2D layers makes possible chemical reactions at the interface between the layers. However, such a possibility has not yet been addressed. Here we demonstrate that HFG bilayer is stable relative to formation of hydrogen fluoride (HF) molecules as a result of chemical reactions between the layers.

In the following, we first give the details of our DFT calculations. In Section IIIA, the results on  structure of functionalized graphene monolayers and HFG bilayer in different symmetric stackings are presented. Then we consider the PES of HFG bilayer and its approximation by the first Fourier harmonics. The characteristics of HFG bilayer associated with relative in-plane motion of the layers are estimated in Section IIIC. In Section IIID, the stability of HFG bilayer relative to formation of HF molecules is addressed. Finally, we discuss PESs for different 2D bilayers.

\section{Computational details}

The DFT calculations are performed using the VASP code \cite{Kresse1996}. The projector augmented-wave method (PAW) \cite{Kresse1999} is applied to describe the interactions of valence electrons with atomic cores. The trigonal unit cell including 4 atoms of each layer (2 C atoms, 1 H atom and 1 F atom in the case of HFG) and having the height of 25~\AA{} is considered under periodic boundary conditions. A dipole correction \cite{Bengtsson1999} is used in the direction perpendicular to the layers to cancel out interactions between periodic images. Integration over the Brillouin zone is performed using the Monkhorst-Pack method \cite{Monkhorst1976} with the $36\times 36 \times 1$ k-point grid. The maximum kinetic energy of plane waves is 600 eV. The convergence threshold of the self-consistent field is $10^{-8}$ eV. The second version of the van der Waals density functional (vdW-DF2) \cite{Lee2010} is used. As follows from the previous comparison with the experimental data \cite{Lebedeva2016a}, this functional in general provides better results for the properties of bilayer graphene, graphite and h-BN related to interlayer interaction, such as shear and bulk moduli, shear mode frequencies, etc. than other functionals corrected for van der Waals interactions (PBE-D2, PBE-D3, PBED3(BJ), PBE-TS and optPBE-vdW). The structure of the monolayers has been also optimized with the exchange-correlation functional of Perdew, Burke and Ernzerhof (PBE) \cite{Perdew1996} for comparison.

To get the structure of the monolayers, both the positions of the atoms and unit cell are changed. To compute the equilibrium interlayer distances for HFG bilayer in different stackings, the positions of the  atoms in the $xy$-plane  parallel to the layers are fixed, while the size of the unit cell and positions of the atoms in the perpendicular $z$-direction are optimized. The geometry optimization is performed till the maximum residual force reaches 0.0003 eV/\AA. The binding energy per carbon atom of one layer is computed as $E_\mathrm{b}=(E_\mathrm{bi}-2E_\mathrm{mono})/n_\mathrm{C}$, where $E_\mathrm{bi}$ is the bilayer energy per unit cell, $E_\mathrm{mono}$ is the monolayer energy per unit cell and $n_\mathrm{C}=2$ is the number of carbon atoms in each layer per unit cell. The 2D polarization $P$ per HFG layer is found as $P = \mu_z/\sigma N$, where $\mu_z$ is the electric dipole moment in the $z$-direction  perpendicular to the layers per unit cell, $\sigma$ is the area of the unit cell in the $xy$-plane parallel to the layers and $N$ is the number of the HFG layers considered. 

To obtain the PES for the coaligned HFG layers, they are placed at the interlayer distance corresponding to the ground-state AA stacking (in which the hydrogen and fluorine atoms of one layer are on top of the similar atoms of the second layer) and then rigidly shifted with respect to each other. The PES for the counteraligned layers is computed at the interlayer distance optimal for the AB1$'$ stacking (in which the hydrogen atoms of one layer are on top of the hydrogen atoms of the second layer and which is the most energetically favourable for the counteraligned layers). The calculations are performed on the grid of $24\times 12$ points, with the step of 0.187~\AA{} and 0.216~\AA{} in the armchair and zigzag directions, respectively. Using the PES symmetry in the zigzag direction, the grid of $24\times 24$ points is finally reconstructed.

The structure and energy of the HF molecule are computed using one $\Gamma$ point in the simulation box with the side of 12~\AA${}={}$1.2~nm and dipole correction in the direction of the bond. 

\section{Results}
\subsection{Structure}
\begin{figure}
   \centering
 \includegraphics{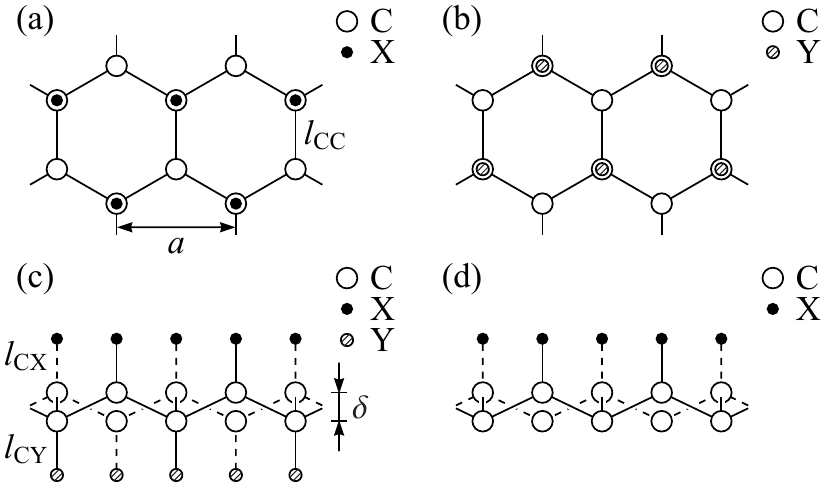}
   \caption{Structures of considered monolayers. (a) Top view of graphane X=H functionalized from two or one sides, hydrofluorinated graphene X=H and fluorographene  X=F functionalized from two or one sides. (b) Bottom view of graphane Y=H, hydrofluorinated graphene Y=F, and fluorographene  Y=F (functionalized from the both sides). (c) Side view of graphane X,Y=H, hydrofluorinated graphene X=H and Y=F, and fluorographene  X,Y=F. (d) Side view of graphane X=H and fluorographene  X=F functionalized from only one side. Lattice constant $a$, bond lengths $l_\mathrm{CC}$, $l_\mathrm{CX}$ and $l_\mathrm{CY}$, and out-of-plane buckling $\delta$ of carbon atoms  are indicated.}
   \label{fig:01}
\end{figure}

First we have calculated the structure of graphane, fluorographene and HFG monolayers and compared the geometrical parameters obtained with literature data. The structure of graphene monolayers hydrogenated or fluorinated from only one side (required to study the stability of HFG bilayer) has been also computed. The chair conformation has been considered for all monolayers since it is the most favourable for graphane \cite{Leenaerts2010, Flores2009, Sofo2007, Samarakoon2009, Samarakoon2011, Antipina2015, Wen2011, Sluiter2003, He2012, Artyukhov2010, Bhattacharya2011},
fluorographene \cite{Charlier1993, Leenaerts2010, Han2010, Samarakoon2011, Antipina2015, Artyukhov2010}, and HFG \cite{Kim2014}. Experimental observations \cite{Mahajan1974, Ebert1974, Parry1974, Touhara1987, Sato2004, Cheng2010} for fluorographite are consistent with this conformation (see also \cite{Mitkin2003} for a review). Different from graphene, where all atoms lie in the same plane (disregarding long-range ripples), carbon atoms of  hydrogenated and fluorinated graphene layers belong to two planes of upper and bottom carbon atoms with the
distance $\delta$~between these planes (see Fig.~\ref{fig:01}a). The distance $\delta$ is referred to here as out-of-plane buckling of carbon atoms.

\begin{table*}
    \caption{Calculated properties of monolayer graphene hydrogenated (X = H) or fluorinated  (X = F) from two or one sides\footnote{Structures of the monolayers are shown in Fig.~\ref{fig:01}}: lattice constant $a$ (in \AA), carbon-carbon length $l_\mathrm{CC}$ (in \AA), carbon-X bond length $l_\mathrm{CX}$ (in \AA), angles $\theta_\mathrm{CCC}$ and $\theta_\mathrm{CCX}$ (in degrees), torsional angles $\theta_\mathrm{CCCC}$ and  $\theta_\mathrm{CCCX}$ (in degrees) and out-of-plane buckling of carbon atoms $\delta$ (in \AA).}
   \renewcommand{\arraystretch}{1.0}
   \setlength{\tabcolsep}{6pt}
    \resizebox{1.0\textwidth}{!}{
        \begin{tabular}{*{10}{c}}
\hline
Approach & $a$ & $l_\mathrm{CC}$  & $l_\mathrm{CX}$  & $\theta_\mathrm{CCC}$  & $\theta_\mathrm{CCX}$ & $\theta_\mathrm{CCCC}$  &$\theta_\mathrm{CCCX}$  &  $\delta$   &   Ref. \\\hline
\multicolumn{10}{c}{Graphane (X = H), two-side functionalization} \\\hline
vdw-DF2 &  2.560  &  1.549  &  1.105 & 111.4   & 107.5  & 54.9  &  62.5 & 0.465  & This  work \\\hline
PBE &  2.541  &  1.537  &  1.110 &  111.5  & 107.4  & 54.7  & 62.7  & 0.459  & This  work \\\hline
PBE &  2.545  &  1.538--1.539  & 1.112 &  111.4  & 107.2--107.9  & ~  & 62.6--62.9  & ~   & \cite{Artyukhov2010} \\\hline
PBE &  2.539  &  1.536  &  1.104 &  111.5  & 107.4  & ~  & ~  & ~  & \cite{Leenaerts2010} \\\hline
PBE &  2.540  &  1.536  &  1.111 &  111.5  & 107.4  & ~  & ~  & 0.458   & \cite{Medeiros2010} \\\hline
PBE &  ~  &  1.526  &  1.110 & 102.8  & 107.5  & ~  & ~  & ~   & \cite{Boukhvalov2008} \\\hline
PBE & 2.54  & 1.54 & 1.11 & 111.5 & 107.4 &~ & ~ &~ & \cite{Singh2011} \\\hline
PBE &  ~  &  1.539 &  1.112   & ~  & ~  & 54.4  & 62.8  & ~  & \cite{Wen2011} \\\hline
PBE &  2.516  &  1.52  &  1.1 & ~  & ~  & ~  & ~  & ~  & \cite{Sofo2007} \\\hline
PBE & 2.55 &  1.54  &  ~  & ~  & ~  & ~  & ~  & ~  & \cite{Samarakoon2009,Samarakoon2011} \\\hline
PBE & 2.54 &  1.537  &  ~  & ~  & ~  & ~  & ~  & ~   & \cite{Flores2009} \\\hline
PBE & ~  &  1.56  &  1.10  & ~  & ~  & ~  & ~  & ~   & \cite{Antipina2015} \\\hline
PBE-D2 & 2.54  &  ~  &  1.107  & ~  & ~  & ~  & ~   & ~ & \cite{Li2012} \\\hline
PW91 & ~  &  1.53 &  1.11 & 111.5 & 107.4  & ~  & ~  & ~  & \cite{Bhattacharya2011} \\\hline
PW91 & 2.504  &  1.537 &  1.110 & ~  & ~  & ~  & ~  & ~   & \cite{He2012} \\\hline
LDA &  2.51  &  1.52  &  1.12 &  112  & 107  & ~  & ~  & 0.45   & \cite{Sahin2009} \\\hline
LDA &  ~  &  1.52  &  1.12 &  111.6  & 107.3  & ~  & ~  & ~   & \cite{Yang2014} \\\hline
LDA &  ~  &  1.53  &  1.09 & ~  & ~  & ~  & ~  & ~   & \cite{Tang2011} \\\hline
exp. &  2.42  &  ~ &  ~  & ~  & ~  & ~  & ~  & ~   & \cite{Elias2009} \\\hline
\multicolumn{10}{c}{Graphane (X = H), one-side functionalization} \\\hline
vdw-DF2 &  2.543  &  1.493  &  1.165 & 116.8   & 100.5  & 34.9 &   72.5 &  0.271   & This  work \\\hline
\multicolumn{10}{c}{Fluorographene (X = F), two-side functionalization} \\\hline
vdw-DF2 &  2.625  &  1.589  &  1.407 &  111.3  & 107.5  &  55.1 &  62.5 & 0.479 &  This  work \\\hline
PBE &  2.609  &  1.583  &  1.382 &  111.0  & 107.9  & 56.2  & 61.9  & 0.488 &  This  work \\\hline
PBE &  2.611  &  1.584  & 1.382 &  110.9  & 108.0  & ~  & 61.9 & ~  &  \cite{Artyukhov2010} \\\hline
PBE &  2.600  &  1.579  &  1.371 &  110.8  & 108.1  & ~  & ~  & ~  &  \cite{Leenaerts2010} \\\hline
PBE &  2.607  &  1.583  &  1.378 &  110.8  & 108.1  & ~  & ~  & 0.490  &  \cite{Medeiros2010} \\\hline
PBE & 2.61  & 1.58 & 1.39 & 111.1 & 107.8 &~ & ~ &~ & \cite{Singh2011} \\\hline
PBE & 2.61 &  ~  &  1.38  & ~  & ~  & ~  & ~  & ~  & \cite{Han2010} \\\hline
PBE & 2.61 &  1.59  &  ~  & ~  & ~  & ~  & ~  & ~  &  \cite{Samarakoon2011} \\\hline
PBE & ~  &  ~  &  1.38  & ~  & ~  & ~  & ~  & ~  & \cite{Antipina2015} \\\hline
PBE-D2 & 2.60  &  1.583  &  1.374  & 110.7 & 108.3  & 56.8  & 61.5 & 0.494 &  \cite{Li2012} \\\hline
LDA &  2.55  &  1.55  &  1.37 &  111  & 108  & ~  & ~  & 0.49  &  \cite{Sahin2011} \\\hline
LDA &  2.553  &  1.552  &  1.37 & 110.7  & 108.2  & ~  & ~  & ~  &  \cite{Charlier1993} \\\hline
LDA &  ~  &  1.55  &  1.37 &  110.7  & 108.3  & ~  & ~  & ~   & \cite{Yang2014} \\\hline
LDA &  2.55  &  1.54  &  1.35   & ~  & ~  & ~  & ~  & ~  & \cite{Takagi2002} \\\hline
LDA &  ~  &  1.56  &  1.35 & ~  & ~  & ~  & ~  & ~  &  \cite{Tang2011} \\\hline
exp. &  2.60--2.61 &  1.58  & 1.36 & 111  & 108  & ~  & ~  & ~  &  \cite{Sato2004} \\\hline
exp. &  2.530$\pm$0.005  &  1.47  & 1.41 & 118.8$\pm$0.5  & ~  & ~  & ~  & ~  &  \cite{Mahajan1974} \\\hline
exp. &  2.54  &  1.54 & 1.39 & ~  & ~  & ~  & ~  & ~  &  \cite{Parry1974} \\\hline
exp. &  2.57  &  1.53 & 1.41 & 109.3  & ~  & ~  & ~  & ~  &  \cite{Touhara1987} \\\hline
exp. &  2.48 & ~  &  ~ & ~  & ~  & ~  & ~  & ~  &  \cite{Nair2010} \\\hline
\multicolumn{10}{c}{Fluorographene (X = F), one-side functionalization} \\\hline
vdw-DF2 &  2.560 &  1.502  &  1.552 &  116.9  & 100.2  & 34.0  & 73.0  & 0.265  & This  work \\\hline\end{tabular}
}
\label{table:monolayer}
\end{table*}

\begin{table*}
    \caption{Properties of hydrofluorinated graphene\footnote{Structure of the monolayer is shown in Fig.~\ref{fig:01}}: lattice constant $a$ (in \AA), carbon-carbon, carbon-hydrogen and carbon-fluorine bond lengths, $l_\mathrm{CC}$, $l_\mathrm{CH}$ and $l_\mathrm{CF}$, respectively, (in \AA), angles $\theta_\mathrm{CCC}$ and $\theta_\mathrm{CCF}$ (in degrees), out-of-plane buckling of carbon atoms $\delta$ (in \AA), and 2D polarization $P$ (in pC/m).}
   \renewcommand{\arraystretch}{1.0}
   \setlength{\tabcolsep}{6pt}
    \resizebox{0.8\textwidth}{!}{
        \begin{tabular}{*{11}{c}}
\hline
Approach & $a$  & $l_\mathrm{CC}$  & $l_\mathrm{CH}$   & $l_\mathrm{CF}$ & $\theta_\mathrm{CCC}$  & $\theta_\mathrm{CCF}$ &  $\delta$  &  $P$   & Ref. \\\hline
vdw-DF2 &  2.594  &  1.570  & 1.103 & 1.408 &  111.4 & 107.5  & 0.472 & 54.39  & This work \\\hline
PBE &  2.576  &  1.560  & 1.106 & 1.386 & 111.2 & 107.6  & 0.473 & 51.37  & This  work \\\hline
PBE &  2.575  &  ~  & 1.106 & 1.386 & ~& ~  & ~  & 47.3  & \cite{Kim2014} \\\hline
PBE-D2 &  2.58  &  ~  & 1.106 & 1.386 & ~& ~  & ~  & 44.2   & \cite{Kim2015} \\\hline
PBE &  2.57  &  1.56  & 1.11 & 1.38  & ~& ~ & 0.48  & ~   &\cite{Cakir2015} \\\hline
PBE &  2.573  &  1.560  & 1.107 & 1.379  & 111.2 & 107.7 & 0.476  &  ~  & \cite{Medeiros2010} \\\hline
PBE &  2.57  &  1.56  & 1.10 & 1.40  & 111.3 & 107.5 & ~  &  ~  & \cite{Singh2011} \\\hline
LDA &  2.54  &  1.54  & 1.11 & 1.38  & 111.1 & 107.8 & 0.47  &  ~   & \cite{Aggoune2016} \\\hline
LDA &  ~  &  1.54  & 1.11 & 1.37  & 111.1 & 107.8 & ~  & ~   & \cite{Yang2014} \\\hline
\end{tabular}
}
\label{table:hf_monolayer}
\end{table*}

The geometrical parameters for graphane and fluorographene monolayers are summarized in Table~\ref{table:monolayer} and for HFG monolayer in Table~\ref{table:hf_monolayer}. As seen from Table~\ref{table:monolayer}, the computed structures agree well with the results of previous first-principles calculations using different functionals and experimental data. The geometrical parameters obtained with the PBE functional are exactly the same as in previous calculations using the similar approach. The account of van der Waals interactions through the vdW-DF2 functional leads to a small (within 2\%) increase  in the lattice constant, carbon-carbon and carbon-fluorine bond lengths and a decrease in the carbon-hydrogen bond length. The out-of-plane buckling $\delta$ in this case increases for graphane, decreases for fluorographene and is almost unchanged for HFG.

The lattice constant of fluorographene is 2.5\% greater than that of graphane (Table~\ref{table:monolayer}) and the lattice constant of HFG monolayer lies in between (Table~\ref{table:hf_monolayer}). These differences are mostly related to changes in the carbon-carbon bond length. The carbon-hydrogen(fluorine) bond lengths are virtually the same in graphane(fluorographene) and HFG monolayer. One-side functionalization leads to a decrease in the lattice constant and carbon-carbon lengths and an increase in the carbon-fluorine and carbon-hydrogen bonds as compared to the monolayers functionalized from the both sides. The out-of-plane buckling $\delta$ is 0.46--0.49~\AA{} for the monolayers functionalized from the both sides. It is reduced almost twice, to $\delta \approx 0.27$ \AA, in the case of one-side functionalization.

The geometrical parameters of HFG bilayer in different symmetric stackings are summarized in Table~\ref{table:hf_bilayer}. As seen from comparison of Tables~\ref{table:hf_monolayer} and \ref{table:hf_bilayer}, the internal structure of the layers is almost unaffected by the interlayer interaction. The changes in the bond lengths, angles and out-of-plane buckling induced by the interlayer interaction do not exceed 1\%. The differences in the structures of two interacting layers lie in the same range. Virtually no change in the internal structure of the layers is observed upon changing the bilayer stacking. For the optimal interlayer distances, the changes in the bond lengths for different stackings are within 0.001 \AA~ (Table~\ref{table:hf_bilayer}). The calculation for the AB2 stacking (PES maxima) at the interlayer distance optimal for the AA stacking (PES minima) demonstrates that the internal structure of the layers is also poorly affected by the changes in the interlayer distance for the distances between equivalent planes of carbon atoms of the top and bottom layers in the range of 5.17--5.37 \AA. As compared to the AB2 stacking at the optimal interlayer distance, the lattice constant and carbon-carbon bond length of the structure with the smaller interlayer distance are decreased by only 0.2\%. 
Similar differences are observed in the geometries of the AA and AB2 stackings at the interlayer distance optimal for AA.
This means that the changes in the internal structure of the layers can be neglected upon relative sliding at the constant interlayer distance.

The computed 2D polarization of HFG monolayer (Table~\ref{table:hf_monolayer}) is in good agreement with the results of previous calculations \cite{Kim2014, Kim2015}. The 2D polarization of HFG bilayer obtained in our paper, however, clearly exceeds the result of Ref.~\cite{Kim2015} but this discrepancy can be attributed to the use of different van der Waals-corrected functionals and other differences in the computational details. As compared to the monolayer, the 2D polarization per layer in HFG bilayer  is enhanced by 1.7\% in the AB1 and AA$'$ stackings and reduced by 0--0.2\% in the other symmetric stackings according to our calculations.

\begin{table*}
    \caption{Properties of different stackings of hydrofluorinated graphene bilayer\footnote{Structure of the monolayer and stackings of the bilayer are shown in Figs.~\ref{fig:01} and \ref{fig:02}, respectively} computed using the vdW-DF2 functional: lattice constant $a$ (in \AA), carbon-carbon, carbon-hydrogen and carbon-fluorine bond lengths, $l_\mathrm{CC}$, $l_\mathrm{CH}$ and $l_\mathrm{CF}$, respectively, (in \AA), angles $\theta_\mathrm{CCC}$ and  $\theta_\mathrm{CCF}$ (in degrees) and out-of-plane buckling of carbon atoms $\delta$ (in \AA), for the top and bottom layer (as indicated in the upper and lower lines, respectively), equilibrium distance between equivalent planes with carbon atoms of the top and bottom layers $d_\text{CC}$ (in \AA), equilibrium distance between the planes with fluorine atoms of the top layer and hydrogen atoms of the bottom layer $d_\text{HF}$ (in \AA), binding energy  per carbon atom of one of the layers $E$ (in meV/atom), relative energy with respect to the AA stacking per carbon atom of one layer $\Delta E$ (in meV/atom), and 2D polarization per layer $P$ (in pC/m).}
   \renewcommand{\arraystretch}{1.0}
   \setlength{\tabcolsep}{6pt}
    \resizebox{1.0\textwidth}{!}{
        \begin{tabular}{*{13}{c}}
\hline
Stacking & $a$ & $l_\mathrm{CC}$  & $l_\mathrm{CH}$  & $l_\mathrm{CF}$   & $\theta_\mathrm{CCC}$  & $\theta_\mathrm{CCF}$ &  $\delta$  & $d_\text{CC}$ & $d_\text{HF}$ & $E$ &  $\Delta E$ & $P$    \\\hline
AA  & \begin{tabular}{@{}c@{}} 2.592 \\ 2.57\footnote{Ref.~\cite{Kim2015}, PBE-D2} \end{tabular} & \begin{tabular}{@{}c@{}} 1.568 \\ 1.570 \end{tabular}   & \begin{tabular}{@{}c@{}} 1.103 \\ 1.099 \end{tabular} & \begin{tabular}{@{}c@{}} 1.419 \\ 1.411 \end{tabular} &  \begin{tabular}{@{}c@{}} 111.51 \\ 111.30 \end{tabular} &  \begin{tabular}{@{}c@{}} 107.35 \\ 107.57  \end{tabular}  & \begin{tabular}{@{}c@{}} 0.468 \\ 0.474 \end{tabular} & \begin{tabular}{@{}c@{}}  5.166\\ 5.172 \end{tabular} & 2.180 &   $-$53.47  & 0 &  \begin{tabular}{@{}c@{}} 54.29, \\  29.97$^\mathrm{b}$ \end{tabular} \\\hline
AB1  & 2.592  & \begin{tabular}{@{}c@{}} 1.568 \\ 1.570 \end{tabular}   & \begin{tabular}{@{}c@{}} 1.103 \\ 1.099 \end{tabular} & \begin{tabular}{@{}c@{}} 1.419 \\ 1.411 \end{tabular} &  \begin{tabular}{@{}c@{}} 111.50 \\ 111.30 \end{tabular} &  \begin{tabular}{@{}c@{}} 107.35 \\ 107.57 \end{tabular}  & \begin{tabular}{@{}c@{}} 0.468 \\ 0.474 \end{tabular} & \begin{tabular}{@{}c@{}}  5.171 \\ 5.178 \end{tabular} & 2.186 & $-$53.06 &  0.42 & 54.32    \\\hline
AB2  & 2.592  & \begin{tabular}{@{}c@{}} 1.568 \\ 1.569 \end{tabular}   & \begin{tabular}{@{}c@{}} 1.103 \\ 1.100 \end{tabular} & \begin{tabular}{@{}c@{}} 1.418 \\ 1.411 \end{tabular} &  \begin{tabular}{@{}c@{}} 111.49 \\ 111.32 \end{tabular} &  \begin{tabular}{@{}c@{}} 107.37 \\ 107.55  \end{tabular}  & \begin{tabular}{@{}c@{}} 0.468 \\ 0.473 \end{tabular} & \begin{tabular}{@{}c@{}}  5.368 \\ 5.373 \end{tabular} & 2.382 & $-$46.19  &  \begin{tabular}{@{}c@{}} 7.28, \\ $16.5^\mathrm{b}$, \\ $8^\mathrm{c}$ \end{tabular}  & 55.29    \\\hline
AA$'$  & 2.592  & \begin{tabular}{@{}c@{}} 1.568 \\ 1.569 \end{tabular}   & \begin{tabular}{@{}c@{}} 1.103 \\ 1.100 \end{tabular} & \begin{tabular}{@{}c@{}} 1.418 \\ 1.411 \end{tabular} &  \begin{tabular}{@{}c@{}} 111.49 \\ 111.32 \end{tabular} &  \begin{tabular}{@{}c@{}} 107.37 \\ 107.55  \end{tabular}  & \begin{tabular}{@{}c@{}} 0.468 \\ 0.473 \end{tabular} & \begin{tabular}{@{}c@{}}  5.367 \\ 5.373 \end{tabular} & 2.382 & $-$46.16  &  \begin{tabular}{@{}c@{}}  7.31, \\ $31.5^\mathrm{b}$, \\ $9^\mathrm{c}$ \end{tabular}  & 55.29    \\\hline
AB1$'$  & 2.592  & \begin{tabular}{@{}c@{}} 1.568 \\ 1.570 \end{tabular}   & \begin{tabular}{@{}c@{}} 1.103 \\ 1.099 \end{tabular} & \begin{tabular}{@{}c@{}} 1.419 \\ 1.411 \end{tabular} &  \begin{tabular}{@{}c@{}} 111.51 \\ 111.30 \end{tabular} &  \begin{tabular}{@{}c@{}} 107.35 \\ 107.57 \end{tabular}  & \begin{tabular}{@{}c@{}} 0.468 \\ 0.474 \end{tabular} & \begin{tabular}{@{}c@{}}  5.170 \\ 5.176 \end{tabular} & 2.184 &  $-$53.40 &  \begin{tabular}{@{}c@{}}  0.070, \\ $0^\mathrm{b}$ \end{tabular}  & 54.33   \\\hline
AB2$'$  & 2.592  & \begin{tabular}{@{}c@{}} 1.568 \\ 1.570 \end{tabular}   & \begin{tabular}{@{}c@{}} 1.103 \\ 1.099 \end{tabular} & \begin{tabular}{@{}c@{}} 1.419 \\ 1.411 \end{tabular} &  \begin{tabular}{@{}c@{}} 111.50 \\ 111.30 \end{tabular} &  \begin{tabular}{@{}c@{}} 107.35 \\ 107.57 \end{tabular}  & \begin{tabular}{@{}c@{}} 0.468 \\ 0.474 \end{tabular} & \begin{tabular}{@{}c@{}}  5.172 \\ 5.178 \end{tabular} & 2.186 & \begin{tabular}{@{}c@{}}$-$53.09, \\ $-$63\footnote{Ref.~\cite{Li2015}, PBE-D2} \end{tabular} & \begin{tabular}{@{}c@{}}  0.38, \\ $0.5^\mathrm{b}$, \\ $-2^\mathrm{c}$  \end{tabular}  &  54.31  \\\hline
AB2\footnote{At the interlayer distance optimal for the AA stacking}  & 2.587  & \begin{tabular}{@{}c@{}} 1.565 \\ 1.567 \end{tabular}   & \begin{tabular}{@{}c@{}} 1.103 \\ 1.100 \end{tabular} & \begin{tabular}{@{}c@{}} 1.419 \\ 1.412 \end{tabular} &  \begin{tabular}{@{}c@{}} 111.47 \\ 111.27 \end{tabular} &  \begin{tabular}{@{}c@{}} 107.39 \\ 107.61  \end{tabular}  & \begin{tabular}{@{}c@{}} 0.468 \\ 0.474 \end{tabular} & \begin{tabular}{@{}c@{}} 5.166\\ 5.173 \end{tabular} & 2.180 & $-$43.94  &  9.53 & 54.11    \\\hline
\end{tabular}
}
\label{table:hf_bilayer}
\end{table*}

\begin{figure}
   \centering
\includegraphics{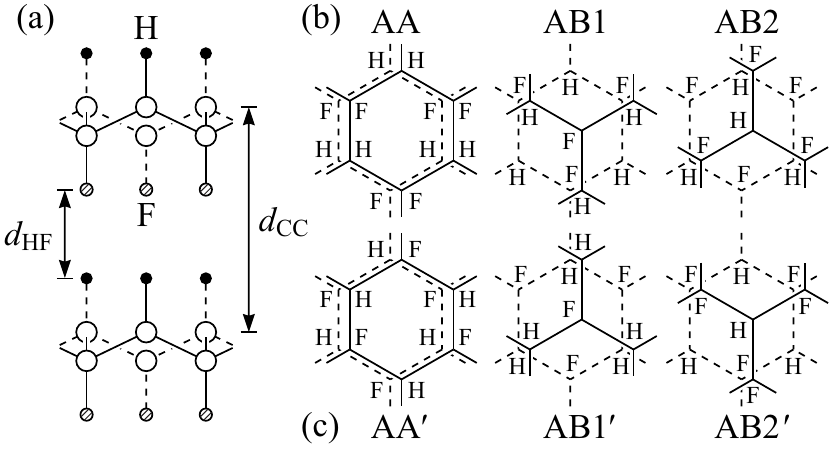}
   \caption{(a) Side view of hydrofluorinated graphene (HFG) bilayer in AA stacking. (b) and (c) Symmetric stackings of HFG bilayer with coaligned and counteraligned layers, respectively. Lower layers are shown by dashed lines.}
   \label{fig:02}
\end{figure}

\subsection{Potential energy surface of hydrofluorinated graphene bilayers}
Our calculations show that the most energetically favourable stacking of HFG bilayer is AA (Table~\ref{table:hf_bilayer}), the stacking in which the layers are coaligned and all hydrogen and fluorine atoms are located on top of the similar atoms of the second layer (Fig.~\ref{fig:02}). Almost the same binding energy (within 0.1 meV per carbon atom of one layer) is obtained also for the AB$1'$ stacking in which the layers are counteraligned and hydrogen atoms are in the ``on-top" positions, while the fluorine atoms are in the centers of the hexagons. The AB1 and AB$2'$ stackings are also close in energy, with the additional cost of only 0.4 meV per carbon atom of one layer compared to the AA stacking. In the AB1 stacking, the layers are coaligned and the fluorine atoms located at the outer side of one layer are on top of the hydrogen atoms located at the outer side of the second layer, while the hydrogen and fluorine atoms between the layers are in the centers of the hexagons. In the AB$2'$ stacking, the layers are counteraligned and the fluorine atoms of one layer are on top of the fluorine atoms of the second layer, while the hydrogen atoms are in the centers of the hexagons. These results are in agreement with Refs.~\cite{Kim2015, Li2015}, where close energies were obtained for the AB$1'$, AB$2'$ and AA stackings (Table~\ref{table:hf_bilayer}).

The PES maxima for the coaligned and counteraligned HFG layers correspond to the AB2 and AA$'$ stackings with the relative energy of about 7.3 meV per carbon atom of one layer (Table~\ref{table:hf_bilayer}). In the AB2 stacking the fluorine and hydrogen atoms located between the layers are in the ``on-top" positions, while the fluorine and hydrogen atoms at the outer sides of the layers are in the middle of the hexagons. In the AA$'$ stacking, the fluorine (hydrogen) atoms of one layer are on top of the hydrogen (fluorine) atoms of the second layer. Large relative energies were also obtained for the AB2 and AA$'$ stackings in Refs.~\cite{Kim2015, Li2015}. The relative energies of these stackings of 8--9 meV per carbon atom of one layer reported in Ref.~\cite{Li2015} are fairly close to our results.  

As can be expected, the spacings between the layers are very close in the AB$1'$, AB$2'$, AB1 and AA stackings. The distances between the equivalent planes of carbon atoms of the layers in these stackings are 5.17--5.18~\AA{} and the distances between the planes of hydrogen and fluorine atoms at the interface are 2.18--2.19~\AA{} (Table~\ref{table:hf_bilayer}). In the AB2 and AA$'$ stackings, these distances are increased by about 0.2 \AA.

\begin{figure*}
   \centering
\includegraphics{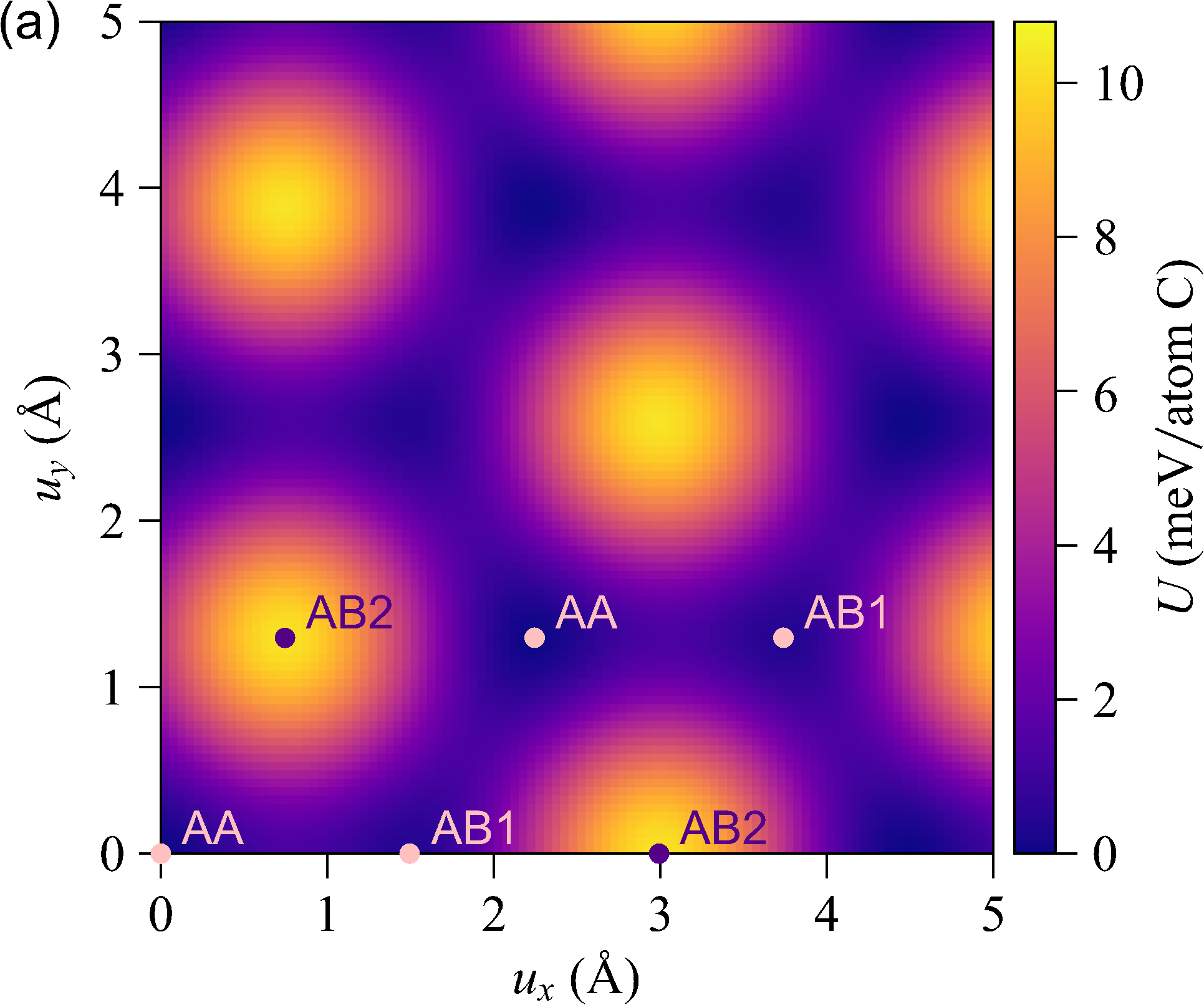}\hfil
\includegraphics{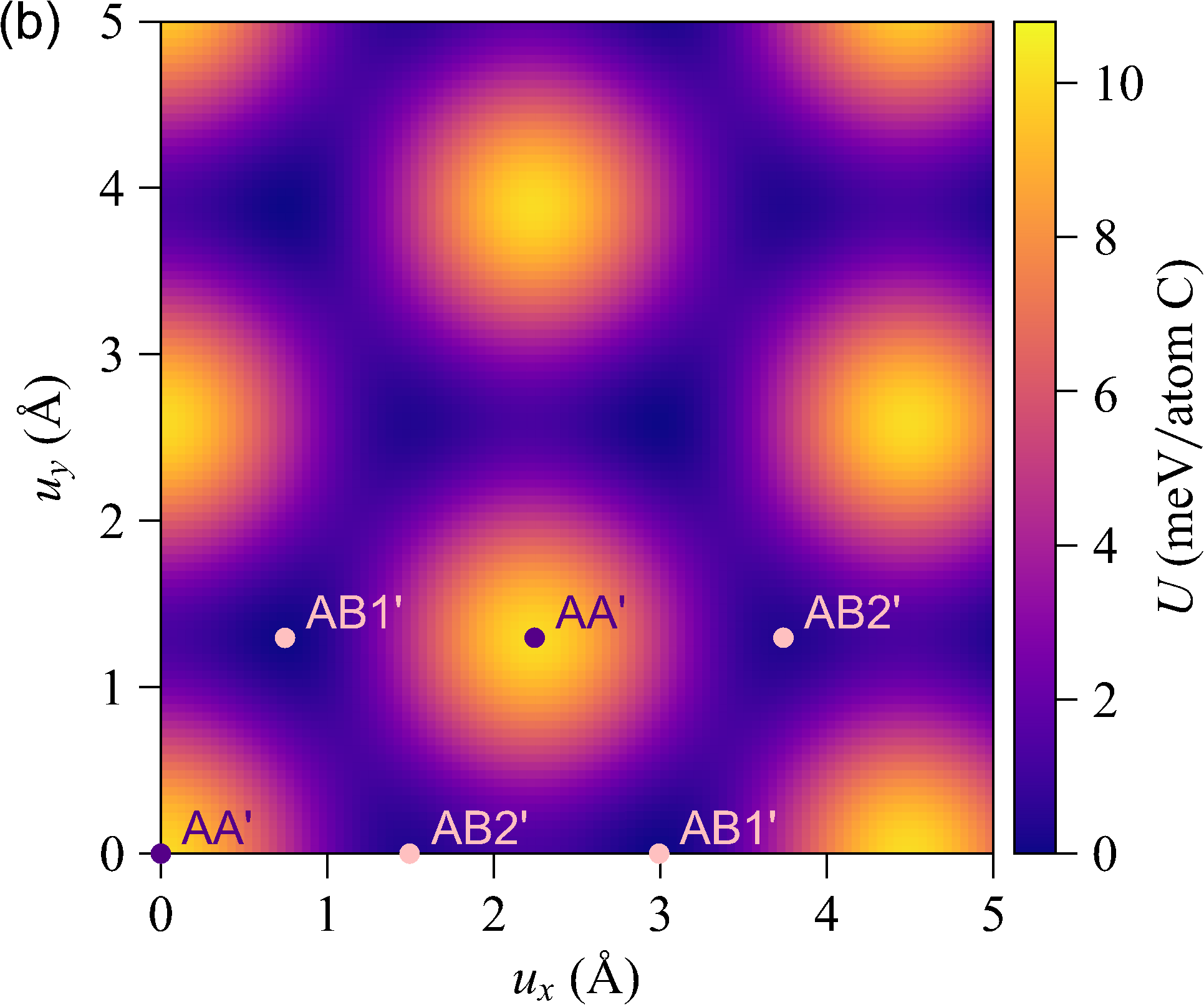}
   \caption{Interlayer interaction energy of hydrofluorinated graphene bilayer $U$ (in meV per carbon atom of one layer) as a function of the relative displacements $u_x$ and $u_y$ (in \AA) of the layers along the armchair and zigzag directions, respectively, approximated according to Eq.~(\ref{eq_U}). Panels (a) and (b) correspond to coaligned and counteraligned layers, respectively. The interlayer distance is constant and equals the optimal one for the AA and AB1$'$ stackings (point $\vec{u}=0$) in the cases of the co- and counteraligned layers, respectively. The energy is also given relative to the AA and AB1$'$ stackings, respectively.}
   \label{fig:03}
\end{figure*}

Let us now derive the expression for the PES of HFG bilayer described by the first Fourier harmonics. We use that the potential energy surface of an atom adsorbed on a 2D hexagonal lattice can be approximated by the first Fourier harmonics as \cite{Verhoeven2004}
\begin{equation} \label{eq_at}
   U_\mathrm{at} = U_1\left[2\cos(k_xu_x)\cos(k_yu_y) + \cos(2k_xu_x) + \frac{3}{2}\right] + U_0,
\end{equation}
where $x$ and $y$ axes are chosen in the armchair and zigzag directions, respectively, $k_x = 2\pi/\!\sqrt{3}a$, $k_y = 2\pi/a$ ($a$ is the lattice constant), $\vec{u}$ describes the relative position of the atom with respect to the  lattice (point $\vec{u} = 0$ corresponds to the case when the atom is located on top of one of the lattice atoms) and parameters $U_1$ and $U_0$ depend on the interlayer distance. The first five terms of the Fourier expansion for the interaction of atoms with a graphene layer and a (111) face of the fcc lattice can be found in Ref.~\cite{Steele1972}. In the case of HFG, we sum up interactions of CF (carbon-fluorine) and CH (carbon-hydrogen) groups with sublattices of CF and CH groups of the second layer, $U = U_\mathrm{FF} + U_\mathrm{HH} + U_\mathrm{HF} + U_\mathrm{FH}$. 

Let us first consider the coaligned layers. In this case,
\begin{align} \label{eq_FF}
   U_\mathrm{FF/HH} = &U_\mathrm{1FF/HH}\Bigg[2\cos(k_xu_x)\cos(k_yu_y) \notag\\
   &+ \cos(2k_xu_x) + \frac{3}{2}\Bigg] + U_\mathrm{0FF/HH},\\
\label{eq_HF}
   U_\mathrm{HF} ={} &U_\mathrm{1HF}\left[2\cos\left(k_x u_x-\frac{2\pi}{3}\right)\cos(k_y u_y)\right. \notag \\
   &+ \left.\cos\left(2k_xu_x - \frac{4\pi}{3}\right) + \frac{3}{2}\right] + U_\mathrm{0HF},\\
%and
 \label{eq_FH}
   U_\mathrm{FH} ={} &U_\mathrm{1FH}\left[2\cos\left(k_xu_x + \frac{2\pi}{3}\right)\cos(k_yu_y)\right. \notag\\
   &+ \left.\cos\left(2k_xu_x + \frac{4\pi}{3}\right) + \frac{3}{2}\right] + U_\mathrm{0FH}.
\end{align}
Here subscript HF corresponds to the interactions of sublattices with the hydrogen and fluorine atoms located at the outer sides of the bilayer and FH to the case when the hydrogen and fluorine atoms are located between the layers. The zero displacement, $\vec{u} = 0$, corresponds to the AA stacking. 

Taking into account Eqs.~(\ref{eq_FF})--(\ref{eq_FH}) for $U = U_\mathrm{FF} + U_\mathrm{HH} + U_\mathrm{HF} + U_\mathrm{FH}$ we finally arrive at the expression
\begin{equation} \label{eq_U}
\begin{split}
   U &= U_\mathrm{A} \bigg(2\cos(k_xu_x)\cos(k_yu_y) + \cos(2k_xu_x) + \frac{3}{2}\bigg)\\
   &+ U_\mathrm{B}\sqrt{3}\bigg(2\sin(k_xu_x)\cos(k_yu_y) - \sin(2k_xu_x)\bigg) + U_\mathrm{C},
   \end{split}
\end{equation}
where for the coaligned layers,
\begin{align} \label{eq_UA}
   U_\mathrm{A} &= U_\mathrm{1FF} + U_\mathrm{1HH} - \frac{1}{2}(U_\mathrm{1HF} + U_\mathrm{1FH}),\\
\label{eq_UB}
   U_\mathrm{B} &= \frac{1}{2}(U_\mathrm{1HF} - U_\mathrm{1FH}),\\
%and
\label{eq_UC}
   U_\mathrm{C} &= U_\mathrm{0} + \frac{9}{4}(U_\mathrm{1HF} + U_\mathrm{1FH}).
\end{align}
Here we introduced the notation $U_\mathrm{0} = U_\mathrm{0FF} + U_\mathrm{0HH} + U_\mathrm{0HF} + U_\mathrm{0FH}$. 

Correspondingly, the energy of the AA stacking is expressed as $E(\mathrm{AA}) = U_\mathrm{C} + 9U_\mathrm{A}/2$ and the energies of the AB1 and AB2 are given by $E(\mathrm{AB1}) = U_\mathrm{C} + 9U_\mathrm{B}/2$ and $E(\mathrm{AB2}) = U_\mathrm{C} - 9U_\mathrm{B}/2$, respectively. Therefore, the parameters of approximation (\ref{eq_U}) can be found as
\begin{align} 
\label{eq_UA1}
   U_\mathrm{A} &= \frac{1}{9}[2E(\mathrm{AA}) - E(\mathrm{AB1}) - E(\mathrm{AB2})],\\
\label{eq_UB1}
   U_\mathrm{B} &= \frac{1}{9}[E(\mathrm{AB1}) - E(\mathrm{AB2})],\\
%and
\label{eq_UC1}
   U_\mathrm{C} &= \frac{1}{2}[E(\mathrm{AB1}) + E(\mathrm{AB2})].
\end{align}

As seen from these equations, $U_\mathrm{A}$ is responsible for the energy of the AA stacking with respect to the average energy of the AB stackings, while $U_\mathrm{B}$ corresponds to the difference between the energies of the AB stackings. The PES corrugation, i.e. the energy difference between the global maximum and minimum, is given by \begin{equation} \label{eq_Umax}
   U_\mathrm{max} = \frac{9}{2}(|U_\mathrm{A}| + |U_\mathrm{B}|).
\end{equation}
%
%,
\begin{table*}
    \caption{Parameters and quality of PES approximation by the first Fourier harmonics for different 2D bilayers at the interlayer distance $d$ (in \AA): parameters $U_\mathrm{A}$, $U_\mathrm{B}$ and $U_\mathrm{C}$ for Eq.~(\ref{eq_U}) per atom of one layer\footnote{Per carbon atom of one layer for HFG bilayer} (in meV/atom), PES corrugation$^\mathrm{b}$ $U_\mathrm{max}$ (in meV/atom), barrier for relative sliding of the layers$^\mathrm{b}$ $U_\mathrm{bar}$ (in meV/atom), standard deviation from the PES obtained in the DFT calculations $\delta U$ (in meV/atom), and relative deviation with respect to the PES corrugation $\delta U/U_\mathrm{max}$ (in \%).}
   \renewcommand{\arraystretch}{1.2}
   \setlength{\tabcolsep}{6pt}
    \resizebox{\textwidth}{!}{
        \begin{tabular}{*{11}{c}}
\hline
Bilayer structure & $d$  & $U_\mathrm{A}$  & $U_\mathrm{B}$ & $U_\mathrm{C}$  &   $U_\mathrm{max}$\footnote{According to the approximation}  &  $U_\mathrm{bar}$    & $\delta U$&$\delta U/U_\mathrm{max}$ & Approach &  Ref. \\\hline
Graphene &  3.25  &  4.24 & 0  & $-$50.59 &  19.08 & 2.12& 0.18 &0.95 & PBE-D2  & \cite{Popov2012} \\\hline
h-BN (coaligned layers) &  3.33  &  3.929 &  0 &  & 17.68 &  1.96 & 0.056 &0.32 & vdW-DF2 & \cite{Lebedev2016} \\\hline
h-BN (counteraligned layers) &  3.33  &  $-$2.098 & 1.408  & 12.35 &   15.77 &  3.57 &  0.014 & 0.09 &vdW-DF2  & \cite{Lebedev2016} \\\hline
Graphene/h-BN heterostructure &  3.33  &  1.662 &   $-$1.082  & &  12.35 &  9.46 & 0.031 & 0.25 &vdW-DF2  & \cite{Lebedev2017} \\\hline
HFG (coaligned layers) &  5.17  & $-$1.182  & $-$1.102 & $-$48.151 &  10.279 & 1.31 & 0.28 & 2.7  & vdW-DF2  & This work \\\hline
HFG (counteraligned layers) &  5.17  &  2.202  & $-$0.037 & $-$53.237 &  10.075 & 1.27 & 0.28  & 2.7 &  vdW-DF2  & This work \\\hline\end{tabular}
}
\label{table:params}
\end{table*}

Here we should use the energies of different stackings at the same interlayer distance. At the interlayer distance optimal  for the ground-state AA stacking, we get $E(\mathrm{AB1}) - E(\mathrm{AA}) = 0.363$~meV/atom and $E(\mathrm{AB2}) - E(\mathrm{AA}) = U_\mathrm{max} = 10.279$~meV/atom.  In this way we obtain the parameters of the approximation listed in Table~\ref{table:params}. The PES described by Eq.~(\ref{eq_U}) with these parameters is shown in Fig.~\ref{fig:03}a. Note that account of relaxation of the internal structure of the HFG layers upon relative sliding at the constant interlayer distance leads to the decrease of the PES corrugation $U_\mathrm{max}$ by only 7\% (see the relative energy of the AB2 stacking at the interlayer distance optimal for the AA stacking in Table~\ref{table:hf_bilayer}).

For the counteraligned layers, 
\begin{align} 
\label{eq_UA2}
   U'_\mathrm{A} &= U_\mathrm{1HF} + U_\mathrm{1FH} - \frac{1}{2}(U_\mathrm{1FF} + U_\mathrm{1HH}),\\
\label{eq_UB2}
   U'_\mathrm{B} &= \frac{1}{2}(U_\mathrm{1HH} - U_\mathrm{1FF}),\\
%and
\label{eq_UC2}
   U'_\mathrm{C} &= U_\mathrm{0} + \frac{9}{4}(U_\mathrm{1FF} + U_\mathrm{1HH}).
\end{align}

Note that from Eqs.~(\ref{eq_UA}), (\ref{eq_UC}), (\ref{eq_UA2}) and (\ref{eq_UC2}), it is seen that at the same interlayer distance, the following condition should be complied for the parameters $U'_\mathrm{A}$,$U_\mathrm{A}$, $U'_\mathrm{C}$ and $U_\mathrm{C}$:
\begin{equation}\label{eq_UCA}
   U'_\mathrm{A} - U_\mathrm{A} = -\frac{2}{3}(U'_\mathrm{C} - U_\mathrm{C}).
\end{equation}

The relations similar to Eqs.~(\ref{eq_UA1})--(\ref{eq_UC1}) hold between $U'_\mathrm{A}$, $U'_\mathrm{B}$ and $U'_\mathrm{C}$ and the energies of the AA$'$, AB$1'$ and AB$2'$ stackings.  At the interlayer distance optimal for the AB1$'$ stacking, $E(\mathrm{AA}') - E(\mathrm{AB1}') = U_\mathrm{max}= 10.075$~meV/atom and $E(\mathrm{AB2}') - E(\mathrm{AB1}') = 0.331$~meV/atom. The values of the parameters that follow from these relative energies are given in Table~\ref{table:params}. The PES described by Eq.~(\ref{eq_U}) with these parameters is shown in Fig.~\ref{fig:03}b.

The standard deviation of expression (\ref{eq_U}) with the parameters from Table~\ref{table:params} from the PES obtained by the DFT calculations (this PES is shown in Fig.~1 of Supplementary Material~\cite{SupplMater}) is 0.28~meV/atom both for co- and counteraligned layers. This corresponds to 2.7\% of the PES corrugation. The maximum deviation of the approximation from the DFT results is 0.39~meV/atom (the full map of deviations is shown in Fig.~2 of Supplementary Material~\cite{SupplMater}). Note that the condition given by Eq.~(\ref{eq_UCA}) is complied well, though the calculations for co- and counteraligned layers are performed at a slightly different interlayer distance (by 0.004 \AA). The difference between the right-hand and left-hand sides of this equation is within 1.6\%. 

It is seen from Table~\ref{table:params} that the values of the parameters $U_\mathrm{A}$ and  $U_\mathrm{B}$ for the HFG layers are very close. The parameter $U'_\mathrm{A}$ is twice greater and $U'_\mathrm{B}$ is small compared to $U_\mathrm{A}$ and  $U_\mathrm{B}$. This means that the term $U_\mathrm{1FH}$, i.e. repulsion of the hydrogen and fluorine atoms between the layers when they are close, dominates over the other pairwise terms (see Eqs.~(\ref{eq_UA}), (\ref{eq_UB}), (\ref{eq_UA2}) and (\ref{eq_UB2})).

\subsection{Properties associated with relative sliding of the layers} 

The PES at a constant interlayer distance can be used to estimate a number of properties associated with relative in-plane motion of the layers that can be measured experimentally \cite{Popov2012, Lebedev2016, Lebedev2017}. Here we 
consider for HFG bilayer the shear mode frequency, shear modulus and barrier for relative rotation of the layers to incommensurate states.

The frequency $f$ of the shear mode $E_{2g}$, in which adjacent layers slide rigidly in the opposite in-plane directions, can be determined from 
 the PES curvature in a given metastable state~\cite{Popov2012, Lebedev2016, Lebedev2017} as
\begin{equation} \label{eq_freq}
   \begin{split}
      f = \frac{1}{2\pi}\sqrt{\frac{1}{\mu}\frac{\partial^2 U}{\partial u_x^2}} = 
      \frac{1}{a}\sqrt{\frac{1}{\mu}U_{\mathrm{eff}}},
   \end{split}
\end{equation}
where  $a$ is the lattice constant, $U_{\mathrm{eff}} = (a/2\pi)^2\, \partial^2 U/\partial u_x^2$ is the second-order derivative of the energy per carbon atom of one layer in energy units  and $\mu$ is the reduced mass. The latter can be computed for the HFG bilayer as $\mu = (2m_\mathrm{C}+m_\mathrm{H}+m_\mathrm{F})/4$, where $m_\mathrm{C}$, $m_\mathrm{F}$ and $m_\mathrm{H}$
are masses of carbon, fluorine and hydrogen atoms, respectively. 

From Eq. (\ref{eq_U}), it follows that the PES curvatures for the AA, AB1 and AB2 stackings correspond to $U_{\mathrm{eff}} (\mathrm{AA}) = -2U_\mathrm{A}$, $U_{\mathrm{eff}} (\mathrm{AB1}) = U_\mathrm{A}-3U_\mathrm{B}$ and $U_{\mathrm{eff}} (\mathrm{AB2}) = U_\mathrm{A}+3U_\mathrm{B}$. Similar expressions hold for the counteraligned layers. From the values of the parameters listed in Table~\ref{table:params}, we thus get that the shear mode frequencies for the AA, AB1, AB1$'$ and AB2$'$ stackings are very close and lie in the range of 17.4--18.5~cm$^{-1}$ (Table~\ref{table:prop}). These values are smaller than those reported for graphene bilayer based on the DFT calculations of 35~cm$^{-1}$~\cite{Lebedeva2011, Lebedeva2012} and 21--34~cm$^{-1}$ (Ref.~\cite{Lebedeva2016a}, depending on the functional used) and experimental studies of $28\pm3$~cm$^{-1}$~\cite{Boschetto2013} and 32~cm$^{-1}$~\cite{Tan2012}. They are also smaller than the DFT results for h-BN bilayer of 33--34~cm$^{-1}$~\cite{Lebedev2016} and 25--47~cm$^{-1}$~\cite{Lebedeva2016a} and graphene/h-BN heterostructure of 37~cm$^{-1}$~\cite{Lebedev2017}. The difference with the data for graphene~\cite{Lebedeva2011, Lebedeva2012, Popov2012, Lebedeva2016a} and h-BN \cite{Lebedev2016, Lebedeva2016a} bilayers can be explained by the smaller PES corrugation (Table~\ref{table:params}) and larger reduced mass for HFG bilayer. As for the graphene/h-BN heterostructure, it has a completely different PES \cite{Lebedev2017}. 

The same PES curvature also determines the shear modulus, which can be estimated as \cite{Lebedev2016}
\begin{equation} \label{eq_C44}
   \begin{split}
      C_{44} = \frac{d}{\sigma}\frac{\partial^2 U}{\partial u_x^2} = 
%      \frac{4\pi^2 d}{a^2 \sigma} U_{\mathrm{eff}}
      \frac{16\pi^2 d}{\sqrt{3}\,a^4} U_{\mathrm{eff}},
   \end{split}
\end{equation}
where $\sigma =\sqrt{3}a^2/4$ is the area per carbon atom in the HFG layer and $d = 5.17$~\AA{} is the interlayer distance. The estimated shear moduli for the AA, AB1, AB1$'$ and AB2$'$ stackings are 3.5--4.0~GPa (Table \ref{table:prop}). The most adequate DFT values of shear moduli reported previously for graphene and h-BN bilayers at the experimental interlayer distance \cite{Lebedeva2016a} lie in the ranges 3.8--4.1~GPa and 4.7--5.6~GPa, respectively. 

When the HFG layers are rotated with respect to each other by an arbitrary angle that does not correspond to a moir\'e pattern, the PES should become smooth, similar to graphene \cite{Popov2012, Lebedeva2010, Lebedeva2011a}. Even in the structures corresponding to moir\'e patterns, the PES corrugation is known to be very small~\cite{Yakobson}. Therefore, the interaction energy in such incommensurate states can be estimated as an average over the PES in the commensurate state given by Eq.~(\ref{eq_U}):
\begin{equation} \label{eq_Urot}
   \begin{split}
      U_{\mathrm{rot}} = \langle U \rangle_{u_x,u_y}  = \frac{3}{2} U_\text{A} + U_\text{C}.
   \end{split}
\end{equation}

\begin{table}
    \caption{Shear mode frequencies $f$ (in cm$^{-1}$), shear moduli $C_{44}$ (in GPa) and barriers $\Delta U_\mathrm{rot}$ (in meV per carbon atom of one layer) for relative rotation of the layers to incommensurate states estimated for different stackings of HFG bilayer corresponding to energy minima.}
   \renewcommand{\arraystretch}{1.2}
   \setlength{\tabcolsep}{12pt}
    \resizebox{0.8\columnwidth}{!}{
        \begin{tabular}{*{9}{c}}
\hline
Stacking & $f$  & $C_{44}$  & $\Delta U_\mathrm{rot}$  \\\hline
AA &  18.53 &  3.96 & 3.55  \\\hline
AB1 &  17.56  &  3.55 & 3.19  \\\hline
AB1$'$ &  18.32  &  3.87 & 3.47  \\\hline
AB2$'$ &  17.43  &  3.50 & 3.14  \\\hline\end{tabular}
}
\label{table:prop}
\end{table}

The barrier $\Delta U_{\mathrm{rot}}$ for relative rotation of the layers to incommensurate states can be find by substracting from $U_{\mathrm{rot}}$ the energy in the minimum. The values of $\Delta U_{\mathrm{rot}}$ estimated for the AA, AB1, AB1$'$ and AB2$'$ stackings lie in the range 3.1--3.6~meV per carbon atom of one layer (Table~\ref{table:prop}). For the same reasons as the shear mode frequency and shear modulus, these barriers are smaller than the previous predictions for graphene bilayer of 4~meV/atom~\cite{Lebedeva2010, Lebedeva2011a} and 5~meV/atom~\cite{Popov2012}, h-BN bilayer of 6.3~meV/atom~\cite{Lebedev2016} and graphene/h-BN heterostructure of 7.4~meV/atom~\cite{Lebedev2017} and 7.0~meV/atom~\cite{Sachs2011}.

\subsection{Stability of hydrofluorinated graphene bilayer}

The recent progress in synthesis of various chemically functionalized graphene layers allows us to propose the possibility of chemical reactions at the interface between the layers with different chemical functionalization. Here we consider the stability of the HFG bilayer relative to decomposition into graphene monolayers hydrogenated or fluorinated  from only one side and HF molecules as a result of chemical reactions between the layers. In such a decomposition, one HF molecule per one unit cell of the HFG bilayer is formed. The computed energy of the HF molecule is $-6.383$~eV. The computed total energies of the HFG bilayer and graphene monolayers hydrogenated or fluorinated from only one side are $-40.591$, $-16.680$ and $-14.701$~eV, respectively, per one unit cell. Thus, our calculations  show that the total energy of the HFG bilayer is lower by 2.827~eV per one unit cell than the total binding energy of the products formed upon bilayer decomposition. That is the HFG bilayer is stable with respect to the considered reaction between the layers. This result, however, does not allow us to exclude the possibility of chemical reactions at the interface between some other chemically functionalized 2D layers.

\section{Discussion and conclusions}

The parameters and deviations of the PES approximation by the first Fourier harmonics for different 2D bilayers are summarized in Table~\ref{table:params} including the present results for HFG bilayers with 2D polarization within one layer and the previous results for a set of 2D bilayers without 2D polarization such as graphene bilayer \cite{Popov2012}, h-BN bilayer \cite{Lebedev2016} and graphene/h-BN heterostructure \cite{Lebedev2017}. Both for graphene and coaligned h-BN layers, the minima of the PES corresponding to the AB stackings are degenerate and $U_\mathrm{B} = 0$. Furthermore, the PES corrugations for these materials given by $U_\mathrm{max} = 9U_\mathrm{A}/2$ (see Eq.~(\ref{eq_Umax})) are  close in magnitude. The very small $|U_\mathrm{B}|$ for HFG bilayer with the counteraligned layers corresponds to the PES with a small energy difference between the AB stackings of only $9|U_\mathrm{B}| = 0.3$~meV/atom. The PES corrugation in the latter case is, however, almost twice smaller than for the graphene and h-BN bilayers (Table~\ref{table:params}).

The type of the PES for counteraligned h-BN layers is similar to that for coaligned HFG layers (Table~\ref{table:params}). In the both cases, the parameter $U_\mathrm{A}$ is negative, which means that the AA stacking is the local minimum and one of the AB stackings is the global maximum (see Eqs.~(\ref{eq_UA1}) and (\ref{eq_UB1})). The close values of $U_\mathrm{A}$ and $U_\mathrm{B}$ for the HFG bilayer correspond to the close energies of the AA and AB1 stackings with the energy difference $9(|U_\mathrm{A}| - |U_\mathrm{B}|)/2 = 0.4$~meV/atom. For the h-BN bilayer, the difference between the energy minima is much more pronounced, 3.1~meV/atom, and the AB1 minimum is very shallow. The PES corrugation is also one and a half greater for the h-BN bilayer compared to the HFG bilayer (Table~\ref{table:params}).

The PES for the graphene/h-BN heterostructure is different from the ones discussed above (Table~\ref{table:params}). Here $U_\mathrm{A}$ is positive and $U_\mathrm{B}$ is comparable in magnitude to $U_\mathrm{A}$. In this case the PES has two inequivalent maxima which correspond to the AA stacking and one of the AB stackings.

The average relative deviation of the approximation by the first Fourier harmonics with respect to the maximal corrugation for the PESs obtained in the DFT calculations is within 1\% and 3\% for the bilayers without and with 2D polarization, respectively (Table~\ref{table:params}). Thus, the hypothesis proposed here that the PES of interlayer interaction in diverse 2D materials can be universally described by the first spatial Fourier harmonics is confirmed in all the cases considered so far. Note that for graphene and h-BN bilayers and graphene/h-BN heterostructure, Fourier expansions up to the third term have been also studied \cite{Zhou2015}. According to these calculations, in the cases of h-BN bilayers and graphene/h-BN heterostructure, the parameters corresponding to the second and third terms are more than an order of magnitude smaller than those for the first term. For bilayer graphene, the difference is by a factor of five. Therefore, the results of Ref.~\cite{Zhou2015} also confirm that the first Fourier harmonics are sufficient to describe the PESs of these materials.

For further confirmation of this hypothesis, the set of considered 2D materials should be extended. It would be now interesting to test transition metal dichalcogenides. DFT calculations of energy profiles for in-plane sliding pathways between the symmetric stackings \cite{Liang2008, Tao2014, Levita2015} and PESs \cite{Liang2008, Levita2015} have been recently performed for MoS$_2$ \cite{Liang2008, Tao2014, Levita2015, Carr2018} MoSe$_2$ \cite{Levita2015}, and MoTe$_2$ \cite{Levita2015} bilayers. The  energy profiles and PESs obtained seem to be qualitatively of the same shape as the PES of HFG bilayer calculated here and the energy profiles and PES for h-BN bilayer \cite{Lebedev2016}. Indeed as long as 2D layers consist of two types of alternating units arranged in the honeycomb lattice, it can be expected that the same Eq.~(\ref{eq_U}) holds for the PES. For MoS$_2$, the PES has been approximated by the first three terms of the Fourier expansion \cite{Carr2018}. According to this approximation, the parameters corresponding to the second and third terms are 6--7 times smaller than for the first one. There is a good chance that the PESs of interlayer interaction for other transition metal dichalcogenides can be also accurately approximated by the first Fourier harmonics. However, these materials are beyond the scope of the present paper and will be considered elsewhere. 

Even though in the present paper we limit ourselves to consideration of the PES of HFG bilayer at the constant interlayer distance, the approximation derived can be useful for modeling of a number of properties and phenomena associated with relative sliding of the layers and involving small changes of the interlayer distance. Let us first discuss the phenomena where the changes of the interlayer distance can be neglected and then the way how the PES approximation can be extended to take into account the dependence on the interlayer distance.

It is reasonable assume that the interlayer distance is constant if the relative displacement takes place close to the PES minima. As examples of such properties, we have estimated the shear mode frequency and shear modulus of HFG bilayer (Table \ref{table:prop}). It is also known from the previous DFT calculations \cite{Zhou2015} for graphene and h-BN bilayers as well as graphene/h-BN heterostructure that the optimal interlayer distance depends on the relative displacement of the layers in the way very similar to the potential energy. This means that the interlayer distance should not change much if the relative displacement of the layers lies far away from the PES maxima. For HFG bilayer, the barrier for relative displacement of the layers between adjacent PES minima is small compared to the PES corrugation, similar to graphene and h-BN bilayers (Table~\ref{table:params}). Thus, the same as for graphene and h-BN bilayers \cite{Zhou2015}, it can be expected that the interlayer distance does not change much along the minimum energy path between adjacent energy minima. In such a case, the PES at the constant interlayer distance can be used to model formation of domain walls between commensurate domains with AB1 and AB2 stackings in the supported bilayer when the size of commensurate domains is much larger than the domain wall width \cite{Alden2013, Lin2013, Yankowitz2014, Popov2011, Lebedeva2016, Lebedev2016, Lebedev2017, Lebedeva2019, Lebedeva2019a, Lebedeva2020} and phenomena related to sliding of a flake on the 2D layer of the same material such as atomic-scale slip-stick motion of the flake attached to a microscope tip \cite{Dienwiebel2004, Dienwiebel2005, Filippov2008} and diffusion of the flake in the commensurate state \cite{Lebedeva2010, Lebedeva2011a}. We have also roughly estimated the barrier to relative rotation of the HFG layers to incommensurate states (Table \ref{table:prop}). 

Our previous DFT calculations \cite{Lebedeva2011} revealed that relative energies of symmetric stackings of graphene bilayer depend on the interlayer distance in the same exponential way. This means that the terms corresponding to the first Fourier harmonics can be multiplied by an exponential factor to describe the PES dependence on the interlayer distance. We, however, leave verification of this fact for HFG bilayer beyond the scope of the present paper. The PES approximation with account of the dependence on the interlayer distance can be useful, for example, for modeling of structure and energetics of moir\'e patterns. 
In the limit of large spatial periods, bilayer superstructure corresponds to domain wall networks in which the size of commensurate domains is much greater than the domain wall width. Such domain wall networks can be described analytically \cite{Popov2011, Lebedeva2016, Lebedev2016, Lebedev2017, Lebedeva2019, Lebedeva2019a, Lebedeva2020}. Atomistic models \cite{Wijk2015, Gargiulo2018, Wijk2014, Leven2016, Argentero2017} are extensively used for simulations of bilayer superstructures in the opposite limit of small spatial periods. The PES approximation by the first Fourier harmonics with account of the dependence on the interlayer distance makes possible development of continuum models \cite{Jung2015, Kumar2015} adequate for studies of intermediate cases.
In particular, such models can be employed to simulate the structures formed upon relative rotation of the layers by the angles of about $1^\circ$, at which superconductivity was discovered for twisted bilayer graphene \cite{Cao2018}.

We have also shown in the present paper that HFG bilayer is stable relative to decomposition into graphene monolayers hydrogenated or fluorinated from only one side and HF molecules as a result of chemical reactions between the layers. However, reactions between other types of functionalized 2D layers cannot be excluded and may require further investigation.

The raw data required to reproduce our findings are available to download from Ref.~\cite{Lebedev2020}.
%https://data.mendeley.com/datasets/tm29mrdnbj/1 (Ref.~\cite{Lebedev2020}).

\hfil
\section*{Acknowledgments}
N.A.P. and S.A.V. acknowledge support by the Belarusian Republican Foundation for Fundamental Research (Grant No.~F20R-301) and Belarusian National Research Program ``Convergence-2020''.
A.M.P. acknowledge support by the Russian Foundation for Basic Research (Grant No. 18-52-00002).

The authors declare no conflict of interest.

\bibliography{rsc}

\begin{thebibliography}{112}%
\makeatletter
\providecommand \@ifxundefined [1]{%
 \@ifx{#1\undefined}
}%
\providecommand \@ifnum [1]{%
 \ifnum #1\expandafter \@firstoftwo
 \else \expandafter \@secondoftwo
 \fi
}%
\providecommand \@ifx [1]{%
 \ifx #1\expandafter \@firstoftwo
 \else \expandafter \@secondoftwo
 \fi
}%
\providecommand \natexlab [1]{#1}%
\providecommand \enquote  [1]{``#1''}%
\providecommand \bibnamefont  [1]{#1}%
\providecommand \bibfnamefont [1]{#1}%
\providecommand \citenamefont [1]{#1}%
\providecommand \href@noop [0]{\@secondoftwo}%
\providecommand \href [0]{\begingroup \@sanitize@url \@href}%
\providecommand \@href[1]{\@@startlink{#1}\@@href}%
\providecommand \@@href[1]{\endgroup#1\@@endlink}%
\providecommand \@sanitize@url [0]{\catcode `\\12\catcode `\$12\catcode
  `\&12\catcode `\#12\catcode `\^12\catcode `\_12\catcode `\%12\relax}%
\providecommand \@@startlink[1]{}%
\providecommand \@@endlink[0]{}%
\providecommand \url  [0]{\begingroup\@sanitize@url \@url }%
\providecommand \@url [1]{\endgroup\@href {#1}{\urlprefix }}%
\providecommand \urlprefix  [0]{URL }%
\providecommand \Eprint [0]{\href }%
\providecommand \doibase [0]{http://dx.doi.org/}%
\providecommand \selectlanguage [0]{\@gobble}%
\providecommand \bibinfo  [0]{\@secondoftwo}%
\providecommand \bibfield  [0]{\@secondoftwo}%
\providecommand \translation [1]{[#1]}%
\providecommand \BibitemOpen [0]{}%
\providecommand \bibitemStop [0]{}%
\providecommand \bibitemNoStop [0]{.\EOS\space}%
\providecommand \EOS [0]{\spacefactor3000\relax}%
\providecommand \BibitemShut  [1]{\csname bibitem#1\endcsname}%
\let\auto@bib@innerbib\@empty
%</preamble>
\bibitem [{\citenamefont {Novoselov}\ \emph {et~al.}(2004)\citenamefont
  {Novoselov}, \citenamefont {Geim}, \citenamefont {Morozov}, \citenamefont
  {Jiang}, \citenamefont {Zhang}, \citenamefont {Dubonos}, \citenamefont
  {Grigorieva},\ and\ \citenamefont {Firsov}}]{Novoselov2004}%
  \BibitemOpen
  \bibfield  {author} {\bibinfo {author} {\bibfnamefont {K.~S.}\ \bibnamefont
  {Novoselov}}, \bibinfo {author} {\bibfnamefont {A.~K.}\ \bibnamefont {Geim}},
  \bibinfo {author} {\bibfnamefont {S.~V.}\ \bibnamefont {Morozov}}, \bibinfo
  {author} {\bibfnamefont {D.}~\bibnamefont {Jiang}}, \bibinfo {author}
  {\bibfnamefont {Y.}~\bibnamefont {Zhang}}, \bibinfo {author} {\bibfnamefont
  {S.~V.}\ \bibnamefont {Dubonos}}, \bibinfo {author} {\bibfnamefont {I.~V.}\
  \bibnamefont {Grigorieva}}, \ and\ \bibinfo {author} {\bibfnamefont {A.~A.}\
  \bibnamefont {Firsov}},\ }\href {\doibase 10.1126/science.1102896} {\bibfield
   {journal} {\bibinfo  {journal} {Science}\ }\textbf {\bibinfo {volume}
  {306}},\ \bibinfo {pages} {666} (\bibinfo {year} {2004})}\BibitemShut
  {NoStop}%
\bibitem [{\citenamefont {Zhang}\ \emph {et~al.}(2009)\citenamefont {Zhang},
  \citenamefont {Tang}, \citenamefont {Girit}, \citenamefont {Hao},
  \citenamefont {Martin}, \citenamefont {Zettl}, \citenamefont {Crommie},
  \citenamefont {Shen},\ and\ \citenamefont {Wang}}]{Zhang2009}%
  \BibitemOpen
  \bibfield  {author} {\bibinfo {author} {\bibfnamefont {Y.}~\bibnamefont
  {Zhang}}, \bibinfo {author} {\bibfnamefont {T.}~\bibnamefont {Tang}},
  \bibinfo {author} {\bibfnamefont {C.}~\bibnamefont {Girit}}, \bibinfo
  {author} {\bibfnamefont {Z.}~\bibnamefont {Hao}}, \bibinfo {author}
  {\bibfnamefont {M.~C.}\ \bibnamefont {Martin}}, \bibinfo {author}
  {\bibfnamefont {A.}~\bibnamefont {Zettl}}, \bibinfo {author} {\bibfnamefont
  {M.~F.}\ \bibnamefont {Crommie}}, \bibinfo {author} {\bibfnamefont {Y.~R.}\
  \bibnamefont {Shen}}, \ and\ \bibinfo {author} {\bibfnamefont
  {F.}~\bibnamefont {Wang}},\ }\href {\doibase 10.1038/nature08105} {\bibfield
  {journal} {\bibinfo  {journal} {Nature}\ }\textbf {\bibinfo {volume} {459}},\
  \bibinfo {pages} {820} (\bibinfo {year} {2009})}\BibitemShut {NoStop}%
\bibitem [{\citenamefont {Cao}\ \emph {et~al.}(2018)\citenamefont {Cao},
  \citenamefont {Fatemi}, \citenamefont {Fang}, \citenamefont {Watanabe},
  \citenamefont {Taniguchi}, \citenamefont {Kaxiras},\ and\ \citenamefont
  {Jarillo-Herrero}}]{Cao2018}%
  \BibitemOpen
  \bibfield  {author} {\bibinfo {author} {\bibfnamefont {Y.}~\bibnamefont
  {Cao}}, \bibinfo {author} {\bibfnamefont {V.}~\bibnamefont {Fatemi}},
  \bibinfo {author} {\bibfnamefont {S.}~\bibnamefont {Fang}}, \bibinfo {author}
  {\bibfnamefont {K.}~\bibnamefont {Watanabe}}, \bibinfo {author}
  {\bibfnamefont {T.}~\bibnamefont {Taniguchi}}, \bibinfo {author}
  {\bibfnamefont {E.}~\bibnamefont {Kaxiras}}, \ and\ \bibinfo {author}
  {\bibfnamefont {P.}~\bibnamefont {Jarillo-Herrero}},\ }\href@noop {}
  {\bibfield  {journal} {\bibinfo  {journal} {Nature}\ }\textbf {\bibinfo
  {volume} {556}},\ \bibinfo {pages} {43} (\bibinfo {year} {2018})}\BibitemShut
  {NoStop}%
\bibitem [{\citenamefont {Popov}\ \emph
  {et~al.}(2011{\natexlab{a}})\citenamefont {Popov}, \citenamefont {Lebedeva},
  \citenamefont {Knizhnik}, \citenamefont {Lozovik},\ and\ \citenamefont
  {Potapkin}}]{Popov2011}%
  \BibitemOpen
  \bibfield  {author} {\bibinfo {author} {\bibfnamefont {A.~M.}\ \bibnamefont
  {Popov}}, \bibinfo {author} {\bibfnamefont {I.~V.}\ \bibnamefont {Lebedeva}},
  \bibinfo {author} {\bibfnamefont {A.~A.}\ \bibnamefont {Knizhnik}}, \bibinfo
  {author} {\bibfnamefont {Y.~E.}\ \bibnamefont {Lozovik}}, \ and\ \bibinfo
  {author} {\bibfnamefont {B.~V.}\ \bibnamefont {Potapkin}},\ }\href {\doibase
  10.1103/PhysRevB.84.045404} {\bibfield  {journal} {\bibinfo  {journal} {Phys.
  Rev. B}\ }\textbf {\bibinfo {volume} {84}},\ \bibinfo {pages} {045404}
  (\bibinfo {year} {2011}{\natexlab{a}})}\BibitemShut {NoStop}%
\bibitem [{\citenamefont {Lebedeva}\ and\ \citenamefont
  {Popov}(2019)}]{Lebedeva2019}%
  \BibitemOpen
  \bibfield  {author} {\bibinfo {author} {\bibfnamefont {I.~V.}\ \bibnamefont
  {Lebedeva}}\ and\ \bibinfo {author} {\bibfnamefont {A.~M.}\ \bibnamefont
  {Popov}},\ }\href {\doibase 10.1103/PhysRevB.99.195448} {\bibfield  {journal}
  {\bibinfo  {journal} {Phys. Rev. B}\ }\textbf {\bibinfo {volume} {99}},\
  \bibinfo {pages} {195448} (\bibinfo {year} {2019})}\BibitemShut {NoStop}%
\bibitem [{\citenamefont {Vaezi}\ \emph {et~al.}(2013)\citenamefont {Vaezi},
  \citenamefont {Liang}, \citenamefont {Ngai}, \citenamefont {Yang},\ and\
  \citenamefont {Kim}}]{Vaezi2013}%
  \BibitemOpen
  \bibfield  {author} {\bibinfo {author} {\bibfnamefont {A.}~\bibnamefont
  {Vaezi}}, \bibinfo {author} {\bibfnamefont {Y.}~\bibnamefont {Liang}},
  \bibinfo {author} {\bibfnamefont {D.~H.}\ \bibnamefont {Ngai}}, \bibinfo
  {author} {\bibfnamefont {L.}~\bibnamefont {Yang}}, \ and\ \bibinfo {author}
  {\bibfnamefont {E.-A.}\ \bibnamefont {Kim}},\ }\href {\doibase
  10.1103/PhysRevX.3.021018} {\bibfield  {journal} {\bibinfo  {journal} {Phys.
  Rev. X}\ }\textbf {\bibinfo {volume} {3}},\ \bibinfo {pages} {021018}
  (\bibinfo {year} {2013})}\BibitemShut {NoStop}%
\bibitem [{\citenamefont {Zhang}\ \emph {et~al.}(2013)\citenamefont {Zhang},
  \citenamefont {MacDonald},\ and\ \citenamefont {Mele}}]{Zhang2013}%
  \BibitemOpen
  \bibfield  {author} {\bibinfo {author} {\bibfnamefont {F.}~\bibnamefont
  {Zhang}}, \bibinfo {author} {\bibfnamefont {A.~H.}\ \bibnamefont
  {MacDonald}}, \ and\ \bibinfo {author} {\bibfnamefont {E.~J.}\ \bibnamefont
  {Mele}},\ }\href {\doibase 10.1073/pnas.1308853110} {\bibfield  {journal}
  {\bibinfo  {journal} {Proc. Natl. Acad. Sci. USA}\ }\textbf {\bibinfo
  {volume} {110}},\ \bibinfo {pages} {10546} (\bibinfo {year}
  {2013})}\BibitemShut {NoStop}%
\bibitem [{\citenamefont {Zheng}\ \emph {et~al.}(2008)\citenamefont {Zheng},
  \citenamefont {Jiang}, \citenamefont {Liu}, \citenamefont {Zhu},
  \citenamefont {Jiang}, \citenamefont {Weng}, \citenamefont {Lu},
  \citenamefont {Wang}, \citenamefont {Xue},\ and\ \citenamefont
  {Peng}}]{Zheng2008}%
  \BibitemOpen
  \bibfield  {author} {\bibinfo {author} {\bibfnamefont {Q.}~\bibnamefont
  {Zheng}}, \bibinfo {author} {\bibfnamefont {B.}~\bibnamefont {Jiang}},
  \bibinfo {author} {\bibfnamefont {S.}~\bibnamefont {Liu}}, \bibinfo {author}
  {\bibfnamefont {J.}~\bibnamefont {Zhu}}, \bibinfo {author} {\bibfnamefont
  {Q.}~\bibnamefont {Jiang}}, \bibinfo {author} {\bibfnamefont
  {Y.}~\bibnamefont {Weng}}, \bibinfo {author} {\bibfnamefont {L.}~\bibnamefont
  {Lu}}, \bibinfo {author} {\bibfnamefont {S.}~\bibnamefont {Wang}}, \bibinfo
  {author} {\bibfnamefont {Q.}~\bibnamefont {Xue}}, \ and\ \bibinfo {author}
  {\bibfnamefont {L.}~\bibnamefont {Peng}},\ }\href {\doibase
  10.1103/PhysRevLett.100.067205} {\bibfield  {journal} {\bibinfo  {journal}
  {Phys. Rev. Lett.}\ }\textbf {\bibinfo {volume} {100}},\ \bibinfo {pages}
  {067205} (\bibinfo {year} {2008})}\BibitemShut {NoStop}%
\bibitem [{\citenamefont {Popov}\ \emph
  {et~al.}(2011{\natexlab{b}})\citenamefont {Popov}, \citenamefont {Lebedeva},
  \citenamefont {Knizhnik}, \citenamefont {Lozovik},\ and\ \citenamefont
  {Potapkin}}]{Popov2011a}%
  \BibitemOpen
  \bibfield  {author} {\bibinfo {author} {\bibfnamefont {A.~M.}\ \bibnamefont
  {Popov}}, \bibinfo {author} {\bibfnamefont {I.~V.}\ \bibnamefont {Lebedeva}},
  \bibinfo {author} {\bibfnamefont {A.~A.}\ \bibnamefont {Knizhnik}}, \bibinfo
  {author} {\bibfnamefont {Y.~E.}\ \bibnamefont {Lozovik}}, \ and\ \bibinfo
  {author} {\bibfnamefont {B.~V.}\ \bibnamefont {Potapkin}},\ }\href {\doibase
  10.1103/PhysRevB.84.245437} {\bibfield  {journal} {\bibinfo  {journal} {Phys.
  Rev. B}\ }\textbf {\bibinfo {volume} {84}},\ \bibinfo {pages} {245437}
  (\bibinfo {year} {2011}{\natexlab{b}})}\BibitemShut {NoStop}%
\bibitem [{\citenamefont {Poklonski}\ \emph {et~al.}(2013)\citenamefont
  {Poklonski}, \citenamefont {Siahlo}, \citenamefont {Vyrko}, \citenamefont
  {Popov}, \citenamefont {Lozovik}, \citenamefont {Lebedeva},\ and\
  \citenamefont {Knizhnik}}]{Poklonski13JCTN141}%
  \BibitemOpen
  \bibfield  {author} {\bibinfo {author} {\bibfnamefont {N.~A.}\ \bibnamefont
  {Poklonski}}, \bibinfo {author} {\bibfnamefont {A.~I.}\ \bibnamefont
  {Siahlo}}, \bibinfo {author} {\bibfnamefont {S.~A.}\ \bibnamefont {Vyrko}},
  \bibinfo {author} {\bibfnamefont {A.~M.}\ \bibnamefont {Popov}}, \bibinfo
  {author} {\bibfnamefont {Y.~E.}\ \bibnamefont {Lozovik}}, \bibinfo {author}
  {\bibfnamefont {I.~V.}\ \bibnamefont {Lebedeva}}, \ and\ \bibinfo {author}
  {\bibfnamefont {A.~A.}\ \bibnamefont {Knizhnik}},\ }\href {\doibase
  doi:10.1166/jctn.2013.2670} {\bibfield  {journal} {\bibinfo  {journal} {J.
  Comput. Theor. Nanosci.}\ }\textbf {\bibinfo {volume} {10}},\ \bibinfo
  {pages} {141} (\bibinfo {year} {2013})}\BibitemShut {NoStop}%
\bibitem [{\citenamefont {Kang}\ \emph {et~al.}(2015)\citenamefont {Kang},
  \citenamefont {Kim},\ and\ \citenamefont {Kwon}}]{Kang2015}%
  \BibitemOpen
  \bibfield  {author} {\bibinfo {author} {\bibfnamefont {J.~W.}\ \bibnamefont
  {Kang}}, \bibinfo {author} {\bibfnamefont {K.-S.}\ \bibnamefont {Kim}}, \
  and\ \bibinfo {author} {\bibfnamefont {O.~K.}\ \bibnamefont {Kwon}},\ }\href
  {\doibase 10.1166/jctn.2015.3740} {\bibfield  {journal} {\bibinfo  {journal}
  {J. Comput. Theor. Nanosci.}\ }\textbf {\bibinfo {volume} {12}},\ \bibinfo
  {pages} {387} (\bibinfo {year} {2015})}\BibitemShut {NoStop}%
\bibitem [{\citenamefont {Kang}\ and\ \citenamefont {Lee}(2016)}]{Kang2016}%
  \BibitemOpen
  \bibfield  {author} {\bibinfo {author} {\bibfnamefont {J.~W.}\ \bibnamefont
  {Kang}}\ and\ \bibinfo {author} {\bibfnamefont {K.~W.}\ \bibnamefont {Lee}},\
  }\href {\doibase 10.1166/jnn.2016.11068} {\bibfield  {journal} {\bibinfo
  {journal} {J. Nanosci. Nanotechnol.}\ }\textbf {\bibinfo {volume} {16}},\
  \bibinfo {pages} {2891} (\bibinfo {year} {2016})}\BibitemShut {NoStop}%
\bibitem [{\citenamefont {Auw\"arter}(2019)}]{Auwarter2019}%
  \BibitemOpen
  \bibfield  {author} {\bibinfo {author} {\bibfnamefont {W.}~\bibnamefont
  {Auw\"arter}},\ }\href {\doibase 10.1016/j.surfrep.2018.10.001} {\bibfield
  {journal} {\bibinfo  {journal} {Surf. Sci. Rep.}\ }\textbf {\bibinfo {volume}
  {74}},\ \bibinfo {pages} {1} (\bibinfo {year} {2019})}\BibitemShut {NoStop}%
\bibitem [{\citenamefont {Elias}\ \emph {et~al.}(2009)\citenamefont {Elias},
  \citenamefont {Nair}, \citenamefont {Mohiuddin}, \citenamefont {Morozov},
  \citenamefont {Blake}, \citenamefont {Halsall}, \citenamefont {Ferrari},
  \citenamefont {Boukhvalov}, \citenamefont {Katsnelson}, \citenamefont
  {Geim},\ and\ \citenamefont {Novoselov}}]{Elias2009}%
  \BibitemOpen
  \bibfield  {author} {\bibinfo {author} {\bibfnamefont {D.~C.}\ \bibnamefont
  {Elias}}, \bibinfo {author} {\bibfnamefont {R.~R.}\ \bibnamefont {Nair}},
  \bibinfo {author} {\bibfnamefont {T.~M.~G.}\ \bibnamefont {Mohiuddin}},
  \bibinfo {author} {\bibfnamefont {S.~V.}\ \bibnamefont {Morozov}}, \bibinfo
  {author} {\bibfnamefont {P.}~\bibnamefont {Blake}}, \bibinfo {author}
  {\bibfnamefont {M.~P.}\ \bibnamefont {Halsall}}, \bibinfo {author}
  {\bibfnamefont {A.~C.}\ \bibnamefont {Ferrari}}, \bibinfo {author}
  {\bibfnamefont {D.~W.}\ \bibnamefont {Boukhvalov}}, \bibinfo {author}
  {\bibfnamefont {M.~I.}\ \bibnamefont {Katsnelson}}, \bibinfo {author}
  {\bibfnamefont {A.~K.}\ \bibnamefont {Geim}}, \ and\ \bibinfo {author}
  {\bibfnamefont {K.~S.}\ \bibnamefont {Novoselov}},\ }\href {\doibase
  10.1126/science.1167130} {\bibfield  {journal} {\bibinfo  {journal}
  {Science}\ }\textbf {\bibinfo {volume} {323}},\ \bibinfo {pages} {610}
  (\bibinfo {year} {2009})}\BibitemShut {NoStop}%
\bibitem [{\citenamefont {Shi}\ \emph {et~al.}(2009)\citenamefont {Shi},
  \citenamefont {Li},\ and\ \citenamefont {Li}}]{Shi2015}%
  \BibitemOpen
  \bibfield  {author} {\bibinfo {author} {\bibfnamefont {Y.}~\bibnamefont
  {Shi}}, \bibinfo {author} {\bibfnamefont {H.}~\bibnamefont {Li}}, \ and\
  \bibinfo {author} {\bibfnamefont {L.-J.}\ \bibnamefont {Li}},\ }\href
  {\doibase 10.1039/C4CS00256C} {\bibfield  {journal} {\bibinfo  {journal}
  {Chem. Soc. Rev.}\ }\textbf {\bibinfo {volume} {44}},\ \bibinfo {pages}
  {2744} (\bibinfo {year} {2009})}\BibitemShut {NoStop}%
\bibitem [{\citenamefont {Churchill}\ and\ \citenamefont
  {Jarillo-Herrero}(2014)}]{Churchill2014}%
  \BibitemOpen
  \bibfield  {author} {\bibinfo {author} {\bibfnamefont {H.~O.~H.}\
  \bibnamefont {Churchill}}\ and\ \bibinfo {author} {\bibfnamefont
  {P.}~\bibnamefont {Jarillo-Herrero}},\ }\href {\doibase
  10.1038/nnano.2014.85} {\bibfield  {journal} {\bibinfo  {journal} {Nat.
  Nanotechnol.}\ }\textbf {\bibinfo {volume} {9}},\ \bibinfo {pages} {330}
  (\bibinfo {year} {2014})}\BibitemShut {NoStop}%
\bibitem [{\citenamefont {Zhang}\ \emph {et~al.}(2015)\citenamefont {Zhang},
  \citenamefont {Yang}, \citenamefont {Gao},\ and\ \citenamefont
  {Yakobson}}]{Zhang2015}%
  \BibitemOpen
  \bibfield  {author} {\bibinfo {author} {\bibfnamefont {Z.}~\bibnamefont
  {Zhang}}, \bibinfo {author} {\bibfnamefont {Y.}~\bibnamefont {Yang}},
  \bibinfo {author} {\bibfnamefont {G.}~\bibnamefont {Gao}}, \ and\ \bibinfo
  {author} {\bibfnamefont {B.~I.}\ \bibnamefont {Yakobson}},\ }\href {\doibase
  10.1002/anie.201505425} {\bibfield  {journal} {\bibinfo  {journal} {Angew.
  Chem. Int. Ed.}\ }\textbf {\bibinfo {volume} {54}},\ \bibinfo {pages} {13022}
  (\bibinfo {year} {2015})}\BibitemShut {NoStop}%
\bibitem [{\citenamefont {Yuhara}\ \emph {et~al.}(2018)\citenamefont {Yuhara},
  \citenamefont {Shimazu}, \citenamefont {Ito}, \citenamefont {Ohta},
  \citenamefont {Kurosawa}, \citenamefont {Nakatake},\ and\ \citenamefont
  {Lay}}]{Yuhara2018}%
  \BibitemOpen
  \bibfield  {author} {\bibinfo {author} {\bibfnamefont {J.}~\bibnamefont
  {Yuhara}}, \bibinfo {author} {\bibfnamefont {H.}~\bibnamefont {Shimazu}},
  \bibinfo {author} {\bibfnamefont {K.}~\bibnamefont {Ito}}, \bibinfo {author}
  {\bibfnamefont {A.}~\bibnamefont {Ohta}}, \bibinfo {author} {\bibfnamefont
  {M.}~\bibnamefont {Kurosawa}}, \bibinfo {author} {\bibfnamefont
  {M.}~\bibnamefont {Nakatake}}, \ and\ \bibinfo {author} {\bibfnamefont
  {G.~L.}\ \bibnamefont {Lay}},\ }\href {\doibase 10.1021/acsnano.8b07006}
  {\bibfield  {journal} {\bibinfo  {journal} {ACS Nano.}\ }\textbf {\bibinfo
  {volume} {12}},\ \bibinfo {pages} {11632} (\bibinfo {year}
  {2018})}\BibitemShut {NoStop}%
\bibitem [{\citenamefont {Siahlo}\ \emph {et~al.}(2018)\citenamefont {Siahlo},
  \citenamefont {Poklonski}, \citenamefont {Lebedev}, \citenamefont {Lebedeva},
  \citenamefont {Popov}, \citenamefont {Vyrko}, \citenamefont {Knizhnik},\ and\
  \citenamefont {Lozovik}}]{Siahlo18PRM036001}%
  \BibitemOpen
  \bibfield  {author} {\bibinfo {author} {\bibfnamefont {A.~I.}\ \bibnamefont
  {Siahlo}}, \bibinfo {author} {\bibfnamefont {N.~A.}\ \bibnamefont
  {Poklonski}}, \bibinfo {author} {\bibfnamefont {A.~V.}\ \bibnamefont
  {Lebedev}}, \bibinfo {author} {\bibfnamefont {I.~V.}\ \bibnamefont
  {Lebedeva}}, \bibinfo {author} {\bibfnamefont {A.~M.}\ \bibnamefont {Popov}},
  \bibinfo {author} {\bibfnamefont {S.~A.}\ \bibnamefont {Vyrko}}, \bibinfo
  {author} {\bibfnamefont {A.~A.}\ \bibnamefont {Knizhnik}}, \ and\ \bibinfo
  {author} {\bibfnamefont {Y.~E.}\ \bibnamefont {Lozovik}},\ }\href {\doibase
  10.1103/PhysRevMaterials.2.036001} {\bibfield  {journal} {\bibinfo  {journal}
  {Phys. Rev. Materials}\ }\textbf {\bibinfo {volume} {2}},\ \bibinfo {pages}
  {036001} (\bibinfo {year} {2018})}\BibitemShut {NoStop}%
\bibitem [{\citenamefont {Geim}\ and\ \citenamefont
  {Grigorieva}(2013)}]{Geim2013}%
  \BibitemOpen
  \bibfield  {author} {\bibinfo {author} {\bibfnamefont {A.}~\bibnamefont
  {Geim}}\ and\ \bibinfo {author} {\bibfnamefont {I.}~\bibnamefont
  {Grigorieva}},\ }\href {\doibase 10.1038/nature12385} {\bibfield  {journal}
  {\bibinfo  {journal} {Nature}\ }\textbf {\bibinfo {volume} {499}},\ \bibinfo
  {pages} {419} (\bibinfo {year} {2013})}\BibitemShut {NoStop}%
\bibitem [{\citenamefont {Ershova}\ \emph {et~al.}(2010)\citenamefont
  {Ershova}, \citenamefont {Lillestolen},\ and\ \citenamefont
  {Bichoutskaia}}]{Ershova2010}%
  \BibitemOpen
  \bibfield  {author} {\bibinfo {author} {\bibfnamefont {O.~V.}\ \bibnamefont
  {Ershova}}, \bibinfo {author} {\bibfnamefont {T.~C.}\ \bibnamefont
  {Lillestolen}}, \ and\ \bibinfo {author} {\bibfnamefont {E.}~\bibnamefont
  {Bichoutskaia}},\ }\href {\doibase 10.1039/C000370K} {\bibfield  {journal}
  {\bibinfo  {journal} {Phys. Chem. Chem. Phys.}\ }\textbf {\bibinfo {volume}
  {12}},\ \bibinfo {pages} {6483} (\bibinfo {year} {2010})}\BibitemShut
  {NoStop}%
\bibitem [{\citenamefont {Lebedeva}\ \emph
  {et~al.}(2011{\natexlab{a}})\citenamefont {Lebedeva}, \citenamefont
  {Knizhnik}, \citenamefont {Popov}, \citenamefont {Lozovik},\ and\
  \citenamefont {Potapkin}}]{Lebedeva2011}%
  \BibitemOpen
  \bibfield  {author} {\bibinfo {author} {\bibfnamefont {I.~V.}\ \bibnamefont
  {Lebedeva}}, \bibinfo {author} {\bibfnamefont {A.~A.}\ \bibnamefont
  {Knizhnik}}, \bibinfo {author} {\bibfnamefont {A.~M.}\ \bibnamefont {Popov}},
  \bibinfo {author} {\bibfnamefont {Y.~E.}\ \bibnamefont {Lozovik}}, \ and\
  \bibinfo {author} {\bibfnamefont {B.~V.}\ \bibnamefont {Potapkin}},\ }\href
  {\doibase 10.1039/c0cp02614j} {\bibfield  {journal} {\bibinfo  {journal}
  {Phys. Chem. Chem. Phys.}\ }\textbf {\bibinfo {volume} {13}},\ \bibinfo
  {pages} {5687} (\bibinfo {year} {2011}{\natexlab{a}})}\BibitemShut {NoStop}%
\bibitem [{\citenamefont {Popov}\ \emph
  {et~al.}(2012{\natexlab{a}})\citenamefont {Popov}, \citenamefont {Lebedeva},
  \citenamefont {Knizhnik}, \citenamefont {Lozovik},\ and\ \citenamefont
  {Potapkin}}]{Popov2012}%
  \BibitemOpen
  \bibfield  {author} {\bibinfo {author} {\bibfnamefont {A.~M.}\ \bibnamefont
  {Popov}}, \bibinfo {author} {\bibfnamefont {I.~V.}\ \bibnamefont {Lebedeva}},
  \bibinfo {author} {\bibfnamefont {A.~A.}\ \bibnamefont {Knizhnik}}, \bibinfo
  {author} {\bibfnamefont {Y.~E.}\ \bibnamefont {Lozovik}}, \ and\ \bibinfo
  {author} {\bibfnamefont {B.~V.}\ \bibnamefont {Potapkin}},\ }\href {\doibase
  10.1016/j.cplett.2012.03.082} {\bibfield  {journal} {\bibinfo  {journal}
  {Chem. Phys. Lett.}\ }\textbf {\bibinfo {volume} {536}},\ \bibinfo {pages}
  {82} (\bibinfo {year} {2012}{\natexlab{a}})}\BibitemShut {NoStop}%
\bibitem [{\citenamefont {Lebedeva}\ \emph {et~al.}(2012)\citenamefont
  {Lebedeva}, \citenamefont {Knizhnik}, \citenamefont {Popov}, \citenamefont
  {Lozovik},\ and\ \citenamefont {Potapkin}}]{Lebedeva2012}%
  \BibitemOpen
  \bibfield  {author} {\bibinfo {author} {\bibfnamefont {I.~V.}\ \bibnamefont
  {Lebedeva}}, \bibinfo {author} {\bibfnamefont {A.~A.}\ \bibnamefont
  {Knizhnik}}, \bibinfo {author} {\bibfnamefont {A.~M.}\ \bibnamefont {Popov}},
  \bibinfo {author} {\bibfnamefont {Y.~E.}\ \bibnamefont {Lozovik}}, \ and\
  \bibinfo {author} {\bibfnamefont {B.~V.}\ \bibnamefont {Potapkin}},\ }\href
  {\doibase 10.1016/j.physe.2011.07.018} {\bibfield  {journal} {\bibinfo
  {journal} {Physica E}\ }\textbf {\bibinfo {volume} {44}},\ \bibinfo {pages}
  {949} (\bibinfo {year} {2012})}\BibitemShut {NoStop}%
\bibitem [{\citenamefont {Reguzzoni}\ \emph {et~al.}(2012)\citenamefont
  {Reguzzoni}, \citenamefont {Fasolino}, \citenamefont {Molinari},\ and\
  \citenamefont {Righi}}]{Reguzzoni2012}%
  \BibitemOpen
  \bibfield  {author} {\bibinfo {author} {\bibfnamefont {M.}~\bibnamefont
  {Reguzzoni}}, \bibinfo {author} {\bibfnamefont {A.}~\bibnamefont {Fasolino}},
  \bibinfo {author} {\bibfnamefont {E.}~\bibnamefont {Molinari}}, \ and\
  \bibinfo {author} {\bibfnamefont {M.~C.}\ \bibnamefont {Righi}},\ }\href
  {\doibase 10.1103/PhysRevB.86.245434} {\bibfield  {journal} {\bibinfo
  {journal} {Phys. Rev. B}\ }\textbf {\bibinfo {volume} {86}},\ \bibinfo
  {pages} {245434} (\bibinfo {year} {2012})}\BibitemShut {NoStop}%
\bibitem [{\citenamefont {Lebedev}\ \emph {et~al.}(2016)\citenamefont
  {Lebedev}, \citenamefont {Lebedeva}, \citenamefont {Knizhnik},\ and\
  \citenamefont {Popov}}]{Lebedev2016}%
  \BibitemOpen
  \bibfield  {author} {\bibinfo {author} {\bibfnamefont {A.~V.}\ \bibnamefont
  {Lebedev}}, \bibinfo {author} {\bibfnamefont {I.~V.}\ \bibnamefont
  {Lebedeva}}, \bibinfo {author} {\bibfnamefont {A.~A.}\ \bibnamefont
  {Knizhnik}}, \ and\ \bibinfo {author} {\bibfnamefont {A.~M.}\ \bibnamefont
  {Popov}},\ }\href {\doibase 10.1039/C5RA20882C} {\bibfield  {journal}
  {\bibinfo  {journal} {RSC Advances}\ }\textbf {\bibinfo {volume} {6}},\
  \bibinfo {pages} {6423} (\bibinfo {year} {2016})}\BibitemShut {NoStop}%
\bibitem [{\citenamefont {Jung}\ \emph {et~al.}(2015)\citenamefont {Jung},
  \citenamefont {DaSilva}, \citenamefont {MacDonald},\ and\ \citenamefont
  {Adam}}]{Jung2015}%
  \BibitemOpen
  \bibfield  {author} {\bibinfo {author} {\bibfnamefont {J.}~\bibnamefont
  {Jung}}, \bibinfo {author} {\bibfnamefont {A.~M.}\ \bibnamefont {DaSilva}},
  \bibinfo {author} {\bibfnamefont {A.~H.}\ \bibnamefont {MacDonald}}, \ and\
  \bibinfo {author} {\bibfnamefont {S.}~\bibnamefont {Adam}},\ }\href {\doibase
  10.1038/ncomms7308} {\bibfield  {journal} {\bibinfo  {journal} {Nat.
  Commun.}\ }\textbf {\bibinfo {volume} {6}},\ \bibinfo {pages} {6308}
  (\bibinfo {year} {2015})}\BibitemShut {NoStop}%
\bibitem [{\citenamefont {Kumar}\ \emph {et~al.}(2015)\citenamefont {Kumar},
  \citenamefont {Er}, \citenamefont {Dong}, \citenamefont {Li},\ and\
  \citenamefont {Shenoy}}]{Kumar2015}%
  \BibitemOpen
  \bibfield  {author} {\bibinfo {author} {\bibfnamefont {H.}~\bibnamefont
  {Kumar}}, \bibinfo {author} {\bibfnamefont {D.}~\bibnamefont {Er}}, \bibinfo
  {author} {\bibfnamefont {L.}~\bibnamefont {Dong}}, \bibinfo {author}
  {\bibfnamefont {J.}~\bibnamefont {Li}}, \ and\ \bibinfo {author}
  {\bibfnamefont {V.~B.}\ \bibnamefont {Shenoy}},\ }\href@noop {} {\bibfield
  {journal} {\bibinfo  {journal} {Sci. Rep.}\ }\textbf {\bibinfo {volume}
  {5}},\ \bibinfo {pages} {10872} (\bibinfo {year} {2015})}\BibitemShut
  {NoStop}%
\bibitem [{\citenamefont {Lebedev}\ \emph {et~al.}(2017)\citenamefont
  {Lebedev}, \citenamefont {Lebedeva}, \citenamefont {Popov},\ and\
  \citenamefont {Knizhnik}}]{Lebedev2017}%
  \BibitemOpen
  \bibfield  {author} {\bibinfo {author} {\bibfnamefont {A.~V.}\ \bibnamefont
  {Lebedev}}, \bibinfo {author} {\bibfnamefont {I.~V.}\ \bibnamefont
  {Lebedeva}}, \bibinfo {author} {\bibfnamefont {A.~M.}\ \bibnamefont {Popov}},
  \ and\ \bibinfo {author} {\bibfnamefont {A.~A.}\ \bibnamefont {Knizhnik}},\
  }\href {\doibase 10.1103/PhysRevB.96.085432} {\bibfield  {journal} {\bibinfo
  {journal} {Phys. Rev. B}\ }\textbf {\bibinfo {volume} {96}},\ \bibinfo
  {pages} {085432} (\bibinfo {year} {2017})}\BibitemShut {NoStop}%
\bibitem [{\citenamefont {Vukovi\'c}\ \emph {et~al.}(2003)\citenamefont
  {Vukovi\'c}, \citenamefont {Damnjanovi\'c},\ and\ \citenamefont
  {Milo\v{s}evi\'c}}]{Vucovic2003}%
  \BibitemOpen
  \bibfield  {author} {\bibinfo {author} {\bibfnamefont {T.}~\bibnamefont
  {Vukovi\'c}}, \bibinfo {author} {\bibfnamefont {M.}~\bibnamefont
  {Damnjanovi\'c}}, \ and\ \bibinfo {author} {\bibfnamefont {I.}~\bibnamefont
  {Milo\v{s}evi\'c}},\ }\href@noop {} {\bibfield  {journal} {\bibinfo
  {journal} {Physica E}\ }\textbf {\bibinfo {volume} {16}},\ \bibinfo {pages}
  {259} (\bibinfo {year} {2003})}\BibitemShut {NoStop}%
\bibitem [{\citenamefont {Belikov}\ \emph {et~al.}(2004)\citenamefont
  {Belikov}, \citenamefont {Lozovik}, \citenamefont {Nikolaev},\ and\
  \citenamefont {Popov}}]{Belikov2004}%
  \BibitemOpen
  \bibfield  {author} {\bibinfo {author} {\bibfnamefont {A.~V.}\ \bibnamefont
  {Belikov}}, \bibinfo {author} {\bibfnamefont {Y.~E.}\ \bibnamefont
  {Lozovik}}, \bibinfo {author} {\bibfnamefont {A.~G.}\ \bibnamefont
  {Nikolaev}}, \ and\ \bibinfo {author} {\bibfnamefont {A.~M.}\ \bibnamefont
  {Popov}},\ }\href@noop {} {\bibfield  {journal} {\bibinfo  {journal} {Chem.
  Phys. Lett.}\ }\textbf {\bibinfo {volume} {385}},\ \bibinfo {pages} {72}
  (\bibinfo {year} {2004})}\BibitemShut {NoStop}%
\bibitem [{\citenamefont {Bichoutskaia}\ \emph {et~al.}(2005)\citenamefont
  {Bichoutskaia}, \citenamefont {Popov}, \citenamefont {El-Barbary},
  \citenamefont {Heggie},\ and\ \citenamefont {Lozovik}}]{Bichoutskaia2005}%
  \BibitemOpen
  \bibfield  {author} {\bibinfo {author} {\bibfnamefont {E.}~\bibnamefont
  {Bichoutskaia}}, \bibinfo {author} {\bibfnamefont {A.~M.}\ \bibnamefont
  {Popov}}, \bibinfo {author} {\bibfnamefont {A.}~\bibnamefont {El-Barbary}},
  \bibinfo {author} {\bibfnamefont {M.~I.}\ \bibnamefont {Heggie}}, \ and\
  \bibinfo {author} {\bibfnamefont {Y.~E.}\ \bibnamefont {Lozovik}},\
  }\href@noop {} {\bibfield  {journal} {\bibinfo  {journal} {Phys. Rev. B}\
  }\textbf {\bibinfo {volume} {71}},\ \bibinfo {pages} {113403} (\bibinfo
  {year} {2005})}\BibitemShut {NoStop}%
\bibitem [{\citenamefont {Bichoutskaia}\ \emph {et~al.}(2009)\citenamefont
  {Bichoutskaia}, \citenamefont {Popov}, \citenamefont {Lozovik}, \citenamefont
  {Ershova}, \citenamefont {Lebedeva},\ and\ \citenamefont
  {Knizhnik}}]{Bichoutskaia2009}%
  \BibitemOpen
  \bibfield  {author} {\bibinfo {author} {\bibfnamefont {E.}~\bibnamefont
  {Bichoutskaia}}, \bibinfo {author} {\bibfnamefont {A.~M.}\ \bibnamefont
  {Popov}}, \bibinfo {author} {\bibfnamefont {Y.~E.}\ \bibnamefont {Lozovik}},
  \bibinfo {author} {\bibfnamefont {O.~V.}\ \bibnamefont {Ershova}}, \bibinfo
  {author} {\bibfnamefont {I.~V.}\ \bibnamefont {Lebedeva}}, \ and\ \bibinfo
  {author} {\bibfnamefont {A.~A.}\ \bibnamefont {Knizhnik}},\ }\href@noop {}
  {\bibfield  {journal} {\bibinfo  {journal} {Phys. Rev. B}\ }\textbf {\bibinfo
  {volume} {80}},\ \bibinfo {pages} {165427} (\bibinfo {year}
  {2009})}\BibitemShut {NoStop}%
\bibitem [{\citenamefont {Popov}\ \emph {et~al.}(2009)\citenamefont {Popov},
  \citenamefont {Lozovik}, \citenamefont {Sobennikov},\ and\ \citenamefont
  {Knizhnik}}]{Popov2009}%
  \BibitemOpen
  \bibfield  {author} {\bibinfo {author} {\bibfnamefont {A.~M.}\ \bibnamefont
  {Popov}}, \bibinfo {author} {\bibfnamefont {Y.~E.}\ \bibnamefont {Lozovik}},
  \bibinfo {author} {\bibfnamefont {A.~S.}\ \bibnamefont {Sobennikov}}, \ and\
  \bibinfo {author} {\bibfnamefont {A.~A.}\ \bibnamefont {Knizhnik}},\ }\href
  {\doibase 10.1134/S1063776109040104} {\bibfield  {journal} {\bibinfo
  {journal} {JETP}\ }\textbf {\bibinfo {volume} {108}},\ \bibinfo {pages} {621}
  (\bibinfo {year} {2009})}\BibitemShut {NoStop}%
\bibitem [{\citenamefont {Popov}\ \emph
  {et~al.}(2012{\natexlab{b}})\citenamefont {Popov}, \citenamefont {Lebedeva},\
  and\ \citenamefont {Knizhnik}}]{Popov2012a}%
  \BibitemOpen
  \bibfield  {author} {\bibinfo {author} {\bibfnamefont {A.~M.}\ \bibnamefont
  {Popov}}, \bibinfo {author} {\bibfnamefont {I.~V.}\ \bibnamefont {Lebedeva}},
  \ and\ \bibinfo {author} {\bibfnamefont {A.~A.}\ \bibnamefont {Knizhnik}},\
  }\href@noop {} {\bibfield  {journal} {\bibinfo  {journal} {Appl. Phys.
  Lett.}\ }\textbf {\bibinfo {volume} {100}},\ \bibinfo {pages} {173101}
  (\bibinfo {year} {2012}{\natexlab{b}})}\BibitemShut {NoStop}%
\bibitem [{\citenamefont {Popov}\ \emph {et~al.}(2013)\citenamefont {Popov},
  \citenamefont {Lebedeva}, \citenamefont {Knizhnik}, \citenamefont {Lozovik},\
  and\ \citenamefont {Potapkin}}]{Popov2013}%
  \BibitemOpen
  \bibfield  {author} {\bibinfo {author} {\bibfnamefont {A.~M.}\ \bibnamefont
  {Popov}}, \bibinfo {author} {\bibfnamefont {I.~V.}\ \bibnamefont {Lebedeva}},
  \bibinfo {author} {\bibfnamefont {A.~A.}\ \bibnamefont {Knizhnik}}, \bibinfo
  {author} {\bibfnamefont {Y.~E.}\ \bibnamefont {Lozovik}}, \ and\ \bibinfo
  {author} {\bibfnamefont {B.~V.}\ \bibnamefont {Potapkin}},\ }\href@noop {}
  {\bibfield  {journal} {\bibinfo  {journal} {J. Chem. Phys.}\ }\textbf
  {\bibinfo {volume} {138}},\ \bibinfo {pages} {024703} (\bibinfo {year}
  {2013})}\BibitemShut {NoStop}%
\bibitem [{\citenamefont {Alden}\ \emph {et~al.}(2013)\citenamefont {Alden},
  \citenamefont {Tsen}, \citenamefont {Huang}, \citenamefont {Hovden},
  \citenamefont {Brown}, \citenamefont {Park}, \citenamefont {Muller},\ and\
  \citenamefont {McEuen}}]{Alden2013}%
  \BibitemOpen
  \bibfield  {author} {\bibinfo {author} {\bibfnamefont {J.~S.}\ \bibnamefont
  {Alden}}, \bibinfo {author} {\bibfnamefont {A.~W.}\ \bibnamefont {Tsen}},
  \bibinfo {author} {\bibfnamefont {P.~Y.}\ \bibnamefont {Huang}}, \bibinfo
  {author} {\bibfnamefont {R.}~\bibnamefont {Hovden}}, \bibinfo {author}
  {\bibfnamefont {L.}~\bibnamefont {Brown}}, \bibinfo {author} {\bibfnamefont
  {J.}~\bibnamefont {Park}}, \bibinfo {author} {\bibfnamefont {D.~A.}\
  \bibnamefont {Muller}}, \ and\ \bibinfo {author} {\bibfnamefont {P.~L.}\
  \bibnamefont {McEuen}},\ }\href {\doibase 10.1073/pnas.1309394110} {\bibfield
   {journal} {\bibinfo  {journal} {Proc. Natl. Acad. Sci. USA}\ }\textbf
  {\bibinfo {volume} {110}},\ \bibinfo {pages} {11256} (\bibinfo {year}
  {2013})}\BibitemShut {NoStop}%
\bibitem [{\citenamefont {Lebedeva}\ \emph {et~al.}(2017)\citenamefont
  {Lebedeva}, \citenamefont {Lebedev}, \citenamefont {Popov},\ and\
  \citenamefont {Knizhnik}}]{Lebedeva2016a}%
  \BibitemOpen
  \bibfield  {author} {\bibinfo {author} {\bibfnamefont {I.~V.}\ \bibnamefont
  {Lebedeva}}, \bibinfo {author} {\bibfnamefont {A.~V.}\ \bibnamefont
  {Lebedev}}, \bibinfo {author} {\bibfnamefont {A.~M.}\ \bibnamefont {Popov}},
  \ and\ \bibinfo {author} {\bibfnamefont {A.~A.}\ \bibnamefont {Knizhnik}},\
  }\href {\doibase 10.1016/j.commatsci.2016.11.011} {\bibfield  {journal}
  {\bibinfo  {journal} {Comput. Mater. Sci.}\ }\textbf {\bibinfo {volume}
  {128}},\ \bibinfo {pages} {45} (\bibinfo {year} {2017})}\BibitemShut
  {NoStop}%
\bibitem [{\citenamefont {Zhou}\ \emph {et~al.}(2015)\citenamefont {Zhou},
  \citenamefont {Han}, \citenamefont {Dai}, \citenamefont {Sun},\ and\
  \citenamefont {Srolovitz}}]{Zhou2015}%
  \BibitemOpen
  \bibfield  {author} {\bibinfo {author} {\bibfnamefont {S.}~\bibnamefont
  {Zhou}}, \bibinfo {author} {\bibfnamefont {J.}~\bibnamefont {Han}}, \bibinfo
  {author} {\bibfnamefont {S.}~\bibnamefont {Dai}}, \bibinfo {author}
  {\bibfnamefont {J.}~\bibnamefont {Sun}}, \ and\ \bibinfo {author}
  {\bibfnamefont {D.~J.}\ \bibnamefont {Srolovitz}},\ }\href {\doibase
  10.1103/PhysRevB.92.155438} {\bibfield  {journal} {\bibinfo  {journal} {Phys.
  Rev. B}\ }\textbf {\bibinfo {volume} {92}},\ \bibinfo {pages} {155438}
  (\bibinfo {year} {2015})}\BibitemShut {NoStop}%
\bibitem [{\citenamefont {Carr}\ \emph {et~al.}(2018)\citenamefont {Carr},
  \citenamefont {Massatt}, \citenamefont {Torrisi}, \citenamefont {Cazeaux},
  \citenamefont {Luskin},\ and\ \citenamefont {Kaxiras}}]{Carr2018}%
  \BibitemOpen
  \bibfield  {author} {\bibinfo {author} {\bibfnamefont {S.}~\bibnamefont
  {Carr}}, \bibinfo {author} {\bibfnamefont {D.}~\bibnamefont {Massatt}},
  \bibinfo {author} {\bibfnamefont {S.~B.}\ \bibnamefont {Torrisi}}, \bibinfo
  {author} {\bibfnamefont {P.}~\bibnamefont {Cazeaux}}, \bibinfo {author}
  {\bibfnamefont {M.}~\bibnamefont {Luskin}}, \ and\ \bibinfo {author}
  {\bibfnamefont {E.}~\bibnamefont {Kaxiras}},\ }\href {\doibase
  10.1103/PhysRevB.98.224102} {\bibfield  {journal} {\bibinfo  {journal} {Phys.
  Rev. B}\ }\textbf {\bibinfo {volume} {98}},\ \bibinfo {pages} {224102}
  (\bibinfo {year} {2018})}\BibitemShut {NoStop}%
\bibitem [{\citenamefont {Rohrer}\ and\ \citenamefont
  {Hyldgaard}(2011)}]{Rohrer2011}%
  \BibitemOpen
  \bibfield  {author} {\bibinfo {author} {\bibfnamefont {J.}~\bibnamefont
  {Rohrer}}\ and\ \bibinfo {author} {\bibfnamefont {P.}~\bibnamefont
  {Hyldgaard}},\ }\href {\doibase 10.1103/PhysRevB.83.165423} {\bibfield
  {journal} {\bibinfo  {journal} {Phys. Rev. B}\ }\textbf {\bibinfo {volume}
  {83}},\ \bibinfo {pages} {165423} (\bibinfo {year} {2011})}\BibitemShut
  {NoStop}%
\bibitem [{\citenamefont {Lu}\ and\ \citenamefont {Feng}(2009)}]{Lu2009}%
  \BibitemOpen
  \bibfield  {author} {\bibinfo {author} {\bibfnamefont {Y.~H.}\ \bibnamefont
  {Lu}}\ and\ \bibinfo {author} {\bibfnamefont {Y.~P.}\ \bibnamefont {Feng}},\
  }\href {\doibase 10.1021/jp9067284} {\bibfield  {journal} {\bibinfo
  {journal} {J. Phys. Chem. C}\ }\textbf {\bibinfo {volume} {113}},\ \bibinfo
  {pages} {20841} (\bibinfo {year} {2009})}\BibitemShut {NoStop}%
\bibitem [{\citenamefont {Li}\ \emph {et~al.}(2012)\citenamefont {Li},
  \citenamefont {Li},\ and\ \citenamefont {Chen}}]{Li2012}%
  \BibitemOpen
  \bibfield  {author} {\bibinfo {author} {\bibfnamefont {Y.}~\bibnamefont
  {Li}}, \bibinfo {author} {\bibfnamefont {F.}~\bibnamefont {Li}}, \ and\
  \bibinfo {author} {\bibfnamefont {Z.}~\bibnamefont {Chen}},\ }\href {\doibase
  10.1021/ja3040416} {\bibfield  {journal} {\bibinfo  {journal} {J. Am. Chem.
  Soc.}\ }\textbf {\bibinfo {volume} {134}},\ \bibinfo {pages} {11269}
  (\bibinfo {year} {2012})}\BibitemShut {NoStop}%
\bibitem [{\citenamefont {Tang}\ and\ \citenamefont {Cao}(2014)}]{Tang2014}%
  \BibitemOpen
  \bibfield  {author} {\bibinfo {author} {\bibfnamefont {S.}~\bibnamefont
  {Tang}}\ and\ \bibinfo {author} {\bibfnamefont {X.}~\bibnamefont {Cao}},\
  }\href {\doibase 10.1039/C4CP03291H} {\bibfield  {journal} {\bibinfo
  {journal} {Phys. Chem. Chem. Phys.}\ }\textbf {\bibinfo {volume} {16}},\
  \bibinfo {pages} {23214} (\bibinfo {year} {2014})}\BibitemShut {NoStop}%
\bibitem [{\citenamefont {Kim}\ \emph {et~al.}(2015)\citenamefont {Kim},
  \citenamefont {Noor-A-Alam},\ and\ \citenamefont {Shin}}]{Kim2015}%
  \BibitemOpen
  \bibfield  {author} {\bibinfo {author} {\bibfnamefont {H.~J.}\ \bibnamefont
  {Kim}}, \bibinfo {author} {\bibfnamefont {M.}~\bibnamefont {Noor-A-Alam}}, \
  and\ \bibinfo {author} {\bibfnamefont {Y.-H.}\ \bibnamefont {Shin}},\ }\href
  {\doibase 10.1063/1.4917215} {\bibfield  {journal} {\bibinfo  {journal} {J.
  Appl. Phys.}\ }\textbf {\bibinfo {volume} {117}},\ \bibinfo {pages} {145304}
  (\bibinfo {year} {2015})}\BibitemShut {NoStop}%
\bibitem [{\citenamefont {Ong}\ \emph {et~al.}(2013)\citenamefont {Ong},
  \citenamefont {Duerloo},\ and\ \citenamefont {Reed}}]{Ong2013}%
  \BibitemOpen
  \bibfield  {author} {\bibinfo {author} {\bibfnamefont {M.~T.}\ \bibnamefont
  {Ong}}, \bibinfo {author} {\bibfnamefont {K.-A.~N.}\ \bibnamefont {Duerloo}},
  \ and\ \bibinfo {author} {\bibfnamefont {E.~J.}\ \bibnamefont {Reed}},\
  }\href {\doibase 10.1021/jp311275} {\bibfield  {journal} {\bibinfo  {journal}
  {J. Phys. Chem. C}\ }\textbf {\bibinfo {volume} {117}},\ \bibinfo {pages}
  {3615} (\bibinfo {year} {2013})}\BibitemShut {NoStop}%
\bibitem [{\citenamefont {Kim}\ \emph {et~al.}(2014)\citenamefont {Kim},
  \citenamefont {Noor-A-Alam}, \citenamefont {Son},\ and\ \citenamefont
  {Shin}}]{Kim2014}%
  \BibitemOpen
  \bibfield  {author} {\bibinfo {author} {\bibfnamefont {H.~J.}\ \bibnamefont
  {Kim}}, \bibinfo {author} {\bibfnamefont {M.}~\bibnamefont {Noor-A-Alam}},
  \bibinfo {author} {\bibfnamefont {J.~Y.}\ \bibnamefont {Son}}, \ and\
  \bibinfo {author} {\bibfnamefont {Y.-H.}\ \bibnamefont {Shin}},\ }\href
  {\doibase https://doi.org/10.1016/j.cplett.2014.04.031} {\bibfield  {journal}
  {\bibinfo  {journal} {Chem. Phys. Lett.}\ }\textbf {\bibinfo {volume}
  {603}},\ \bibinfo {pages} {62 } (\bibinfo {year} {2014})}\BibitemShut
  {NoStop}%
\bibitem [{\citenamefont {\c{C}ak\i{}r}\ and\ \citenamefont
  {Peeters}(2015)}]{Cakir2015}%
  \BibitemOpen
  \bibfield  {author} {\bibinfo {author} {\bibfnamefont {D.}~\bibnamefont
  {\c{C}ak\i{}r}}\ and\ \bibinfo {author} {\bibfnamefont {F.~M.}\ \bibnamefont
  {Peeters}},\ }\href {\doibase 10.1039/C5CP04438C} {\bibfield  {journal}
  {\bibinfo  {journal} {Phys. Chem. Chem. Phys.}\ }\textbf {\bibinfo {volume}
  {17}},\ \bibinfo {pages} {27636} (\bibinfo {year} {2015})}\BibitemShut
  {NoStop}%
\bibitem [{\citenamefont {Sofer}\ \emph {et~al.}(2015)\citenamefont {Sofer},
  \citenamefont {\u{S}imek}, \citenamefont {Maz\'{a}nek}, \citenamefont
  {\u{S}embera}, \citenamefont {Janou\u{s}ek},\ and\ \citenamefont
  {Pumera}}]{Sofer2015}%
  \BibitemOpen
  \bibfield  {author} {\bibinfo {author} {\bibfnamefont {Z.}~\bibnamefont
  {Sofer}}, \bibinfo {author} {\bibfnamefont {P.}~\bibnamefont {\u{S}imek}},
  \bibinfo {author} {\bibfnamefont {V.}~\bibnamefont {Maz\'{a}nek}}, \bibinfo
  {author} {\bibfnamefont {F.}~\bibnamefont {\u{S}embera}}, \bibinfo {author}
  {\bibfnamefont {Z.}~\bibnamefont {Janou\u{s}ek}}, \ and\ \bibinfo {author}
  {\bibfnamefont {M.}~\bibnamefont {Pumera}},\ }\href {\doibase
  10.1039/C4CC08844A} {\bibfield  {journal} {\bibinfo  {journal} {Chem.
  Commun.}\ }\textbf {\bibinfo {volume} {51}},\ \bibinfo {pages} {5633}
  (\bibinfo {year} {2015})}\BibitemShut {NoStop}%
\bibitem [{\citenamefont {Medeiros}\ \emph {et~al.}(2010)\citenamefont
  {Medeiros}, \citenamefont {Mascarenhas}, \citenamefont {de~Brito~Mota},\ and\
  \citenamefont {de~Castilho}}]{Medeiros2010}%
  \BibitemOpen
  \bibfield  {author} {\bibinfo {author} {\bibfnamefont {P.~V.~C.}\
  \bibnamefont {Medeiros}}, \bibinfo {author} {\bibfnamefont {A.~J.~S.}\
  \bibnamefont {Mascarenhas}}, \bibinfo {author} {\bibfnamefont
  {F.}~\bibnamefont {de~Brito~Mota}}, \ and\ \bibinfo {author} {\bibfnamefont
  {C.~M.~C.}\ \bibnamefont {de~Castilho}},\ }\href@noop {} {\bibfield
  {journal} {\bibinfo  {journal} {Nanotechnology}\ }\textbf {\bibinfo {volume}
  {21}},\ \bibinfo {pages} {485701} (\bibinfo {year} {2010})}\BibitemShut
  {NoStop}%
\bibitem [{\citenamefont {Singh}\ and\ \citenamefont
  {Bester}(2011)}]{Singh2011}%
  \BibitemOpen
  \bibfield  {author} {\bibinfo {author} {\bibfnamefont {R.}~\bibnamefont
  {Singh}}\ and\ \bibinfo {author} {\bibfnamefont {G.}~\bibnamefont {Bester}},\
  }\href {\doibase 10.1103/PhysRevB.84.155427} {\bibfield  {journal} {\bibinfo
  {journal} {Phys. Rev. B}\ }\textbf {\bibinfo {volume} {84}},\ \bibinfo
  {pages} {155427} (\bibinfo {year} {2011})}\BibitemShut {NoStop}%
\bibitem [{\citenamefont {Aggoune}\ \emph {et~al.}(2016)\citenamefont
  {Aggoune}, \citenamefont {Rezouali},\ and\ \citenamefont
  {Belkhir}}]{Aggoune2016}%
  \BibitemOpen
  \bibfield  {author} {\bibinfo {author} {\bibfnamefont {W.}~\bibnamefont
  {Aggoune}}, \bibinfo {author} {\bibfnamefont {K.}~\bibnamefont {Rezouali}}, \
  and\ \bibinfo {author} {\bibfnamefont {M.~A.}\ \bibnamefont {Belkhir}},\
  }\href {\doibase 10.1002/pssb.201552431} {\bibfield  {journal} {\bibinfo
  {journal} {Phys. Status Solidi B}\ }\textbf {\bibinfo {volume} {253}},\
  \bibinfo {pages} {712} (\bibinfo {year} {2016})}\BibitemShut {NoStop}%
\bibitem [{\citenamefont {Li}\ and\ \citenamefont {Li}(2015)}]{Li2015}%
  \BibitemOpen
  \bibfield  {author} {\bibinfo {author} {\bibfnamefont {F.}~\bibnamefont
  {Li}}\ and\ \bibinfo {author} {\bibfnamefont {Y.}~\bibnamefont {Li}},\ }\href
  {\doibase 10.1039/C5TC00013K} {\bibfield  {journal} {\bibinfo  {journal} {J.
  Mater. Chem. C}\ }\textbf {\bibinfo {volume} {3}},\ \bibinfo {pages} {3416}
  (\bibinfo {year} {2015})}\BibitemShut {NoStop}%
\bibitem [{\citenamefont {Kresse}\ and\ \citenamefont
  {Furthm\"{u}ller}(1996)}]{Kresse1996}%
  \BibitemOpen
  \bibfield  {author} {\bibinfo {author} {\bibfnamefont {G.}~\bibnamefont
  {Kresse}}\ and\ \bibinfo {author} {\bibfnamefont {J.}~\bibnamefont
  {Furthm\"{u}ller}},\ }\href {\doibase 10.1103/PhysRevB.54.11169} {\bibfield
  {journal} {\bibinfo  {journal} {Phys. Rev. B}\ }\textbf {\bibinfo {volume}
  {54}},\ \bibinfo {pages} {11169} (\bibinfo {year} {1996})}\BibitemShut
  {NoStop}%
\bibitem [{\citenamefont {Kresse}\ and\ \citenamefont
  {Joubert}(1999)}]{Kresse1999}%
  \BibitemOpen
  \bibfield  {author} {\bibinfo {author} {\bibfnamefont {G.}~\bibnamefont
  {Kresse}}\ and\ \bibinfo {author} {\bibfnamefont {D.}~\bibnamefont
  {Joubert}},\ }\href {\doibase 10.1103/PhysRevB.59.1758} {\bibfield  {journal}
  {\bibinfo  {journal} {Phys. Rev. B}\ }\textbf {\bibinfo {volume} {59}},\
  \bibinfo {pages} {1758} (\bibinfo {year} {1999})}\BibitemShut {NoStop}%
\bibitem [{\citenamefont {Bengtsson}(1999)}]{Bengtsson1999}%
  \BibitemOpen
  \bibfield  {author} {\bibinfo {author} {\bibfnamefont {L.}~\bibnamefont
  {Bengtsson}},\ }\href {\doibase 10.1103/PhysRevB.59.12301} {\bibfield
  {journal} {\bibinfo  {journal} {Phys. Rev. B}\ }\textbf {\bibinfo {volume}
  {59}},\ \bibinfo {pages} {12301} (\bibinfo {year} {1999})}\BibitemShut
  {NoStop}%
\bibitem [{\citenamefont {Monkhorst}\ and\ \citenamefont
  {Pack}(1976)}]{Monkhorst1976}%
  \BibitemOpen
  \bibfield  {author} {\bibinfo {author} {\bibfnamefont {H.~J.}\ \bibnamefont
  {Monkhorst}}\ and\ \bibinfo {author} {\bibfnamefont {J.~D.}\ \bibnamefont
  {Pack}},\ }\href {\doibase 10.1103/PhysRevB.13.5188} {\bibfield  {journal}
  {\bibinfo  {journal} {Phys. Rev. B}\ }\textbf {\bibinfo {volume} {13}},\
  \bibinfo {pages} {5188} (\bibinfo {year} {1976})}\BibitemShut {NoStop}%
\bibitem [{\citenamefont {Lee}\ \emph {et~al.}(2010)\citenamefont {Lee},
  \citenamefont {Murray}, \citenamefont {Kong}, \citenamefont {Lundqvist},\
  and\ \citenamefont {Langreth}}]{Lee2010}%
  \BibitemOpen
  \bibfield  {author} {\bibinfo {author} {\bibfnamefont {K.}~\bibnamefont
  {Lee}}, \bibinfo {author} {\bibfnamefont {{\'E}.~D.}\ \bibnamefont {Murray}},
  \bibinfo {author} {\bibfnamefont {L.}~\bibnamefont {Kong}}, \bibinfo {author}
  {\bibfnamefont {B.~I.}\ \bibnamefont {Lundqvist}}, \ and\ \bibinfo {author}
  {\bibfnamefont {D.~C.}\ \bibnamefont {Langreth}},\ }\href {\doibase
  10.1103/PhysRevB.82.081101} {\bibfield  {journal} {\bibinfo  {journal} {Phys.
  Rev. B}\ }\textbf {\bibinfo {volume} {82}},\ \bibinfo {pages} {081101(R)}
  (\bibinfo {year} {2010})}\BibitemShut {NoStop}%
\bibitem [{\citenamefont {Perdew}\ \emph {et~al.}(1996)\citenamefont {Perdew},
  \citenamefont {Burke},\ and\ \citenamefont {Ernzerhof}}]{Perdew1996}%
  \BibitemOpen
  \bibfield  {author} {\bibinfo {author} {\bibfnamefont {J.~P.}\ \bibnamefont
  {Perdew}}, \bibinfo {author} {\bibfnamefont {K.}~\bibnamefont {Burke}}, \
  and\ \bibinfo {author} {\bibfnamefont {M.}~\bibnamefont {Ernzerhof}},\ }\href
  {\doibase 10.1103/PhysRevLett.77.3865} {\bibfield  {journal} {\bibinfo
  {journal} {Phys. Rev. Lett.}\ }\textbf {\bibinfo {volume} {77}},\ \bibinfo
  {pages} {3865} (\bibinfo {year} {1996})}\BibitemShut {NoStop}%
\bibitem [{\citenamefont {Leenaerts}\ \emph {et~al.}(2010)\citenamefont
  {Leenaerts}, \citenamefont {Peelaers}, \citenamefont {Hern\'andez-Nieves},
  \citenamefont {Partoens},\ and\ \citenamefont {Peeters}}]{Leenaerts2010}%
  \BibitemOpen
  \bibfield  {author} {\bibinfo {author} {\bibfnamefont {O.}~\bibnamefont
  {Leenaerts}}, \bibinfo {author} {\bibfnamefont {H.}~\bibnamefont {Peelaers}},
  \bibinfo {author} {\bibfnamefont {A.~D.}\ \bibnamefont {Hern\'andez-Nieves}},
  \bibinfo {author} {\bibfnamefont {B.}~\bibnamefont {Partoens}}, \ and\
  \bibinfo {author} {\bibfnamefont {F.~M.}\ \bibnamefont {Peeters}},\ }\href
  {\doibase 10.1103/PhysRevB.82.195436} {\bibfield  {journal} {\bibinfo
  {journal} {Phys. Rev. B}\ }\textbf {\bibinfo {volume} {82}},\ \bibinfo
  {pages} {195436} (\bibinfo {year} {2010})}\BibitemShut {NoStop}%
\bibitem [{\citenamefont {Flores}\ \emph {et~al.}(2009)\citenamefont {Flores},
  \citenamefont {Autreto}, \citenamefont {Legoas},\ and\ \citenamefont
  {Galvao}}]{Flores2009}%
  \BibitemOpen
  \bibfield  {author} {\bibinfo {author} {\bibfnamefont {M.~Z.~S.}\
  \bibnamefont {Flores}}, \bibinfo {author} {\bibfnamefont {P.~A.~S.}\
  \bibnamefont {Autreto}}, \bibinfo {author} {\bibfnamefont {S.~B.}\
  \bibnamefont {Legoas}}, \ and\ \bibinfo {author} {\bibfnamefont {D.~S.}\
  \bibnamefont {Galvao}},\ }\href@noop {} {\bibfield  {journal} {\bibinfo
  {journal} {Nanotechnology}\ }\textbf {\bibinfo {volume} {20}},\ \bibinfo
  {pages} {465704} (\bibinfo {year} {2009})}\BibitemShut {NoStop}%
\bibitem [{\citenamefont {Sofo}\ \emph {et~al.}(2007)\citenamefont {Sofo},
  \citenamefont {Chaudhari},\ and\ \citenamefont {Barber}}]{Sofo2007}%
  \BibitemOpen
  \bibfield  {author} {\bibinfo {author} {\bibfnamefont {J.~O.}\ \bibnamefont
  {Sofo}}, \bibinfo {author} {\bibfnamefont {A.~S.}\ \bibnamefont {Chaudhari}},
  \ and\ \bibinfo {author} {\bibfnamefont {G.~D.}\ \bibnamefont {Barber}},\
  }\href {\doibase 10.1103/PhysRevB.75.153401} {\bibfield  {journal} {\bibinfo
  {journal} {Phys. Rev. B}\ }\textbf {\bibinfo {volume} {75}},\ \bibinfo
  {pages} {153401} (\bibinfo {year} {2007})}\BibitemShut {NoStop}%
\bibitem [{\citenamefont {Samarakoon}\ and\ \citenamefont
  {Wang}(2009)}]{Samarakoon2009}%
  \BibitemOpen
  \bibfield  {author} {\bibinfo {author} {\bibfnamefont {D.~K.}\ \bibnamefont
  {Samarakoon}}\ and\ \bibinfo {author} {\bibfnamefont {X.-Q.}\ \bibnamefont
  {Wang}},\ }\href {\doibase 10.1021/nn901317d} {\bibfield  {journal} {\bibinfo
   {journal} {ACS Nano}\ }\textbf {\bibinfo {volume} {3}},\ \bibinfo {pages}
  {4017} (\bibinfo {year} {2009})}\BibitemShut {NoStop}%
\bibitem [{\citenamefont {Samarakoon}\ \emph {et~al.}(2011)\citenamefont
  {Samarakoon}, \citenamefont {Chen}, \citenamefont {Nicolas},\ and\
  \citenamefont {Wang}}]{Samarakoon2011}%
  \BibitemOpen
  \bibfield  {author} {\bibinfo {author} {\bibfnamefont {D.~K.}\ \bibnamefont
  {Samarakoon}}, \bibinfo {author} {\bibfnamefont {Z.}~\bibnamefont {Chen}},
  \bibinfo {author} {\bibfnamefont {C.}~\bibnamefont {Nicolas}}, \ and\
  \bibinfo {author} {\bibfnamefont {X.-Q.}\ \bibnamefont {Wang}},\ }\href
  {\doibase 10.1002/smll.201002058} {\bibfield  {journal} {\bibinfo  {journal}
  {Small}\ }\textbf {\bibinfo {volume} {7}},\ \bibinfo {pages} {965} (\bibinfo
  {year} {2011})}\BibitemShut {NoStop}%
\bibitem [{\citenamefont {Antipina}\ and\ \citenamefont
  {Sorokin}(2015)}]{Antipina2015}%
  \BibitemOpen
  \bibfield  {author} {\bibinfo {author} {\bibfnamefont {L.~Y.}\ \bibnamefont
  {Antipina}}\ and\ \bibinfo {author} {\bibfnamefont {P.~B.}\ \bibnamefont
  {Sorokin}},\ }\href {\doibase 10.1021/jp510390b} {\bibfield  {journal}
  {\bibinfo  {journal} {J. Phys. Chem. C}\ }\textbf {\bibinfo {volume} {119}},\
  \bibinfo {pages} {2828} (\bibinfo {year} {2015})}\BibitemShut {NoStop}%
\bibitem [{\citenamefont {Wen}\ \emph {et~al.}(2011)\citenamefont {Wen},
  \citenamefont {Hand}, \citenamefont {Labet}, \citenamefont {Yang},
  \citenamefont {Hoffmann}, \citenamefont {Ashcroft}, \citenamefont {Oganov},\
  and\ \citenamefont {Lyakhov}}]{Wen2011}%
  \BibitemOpen
  \bibfield  {author} {\bibinfo {author} {\bibfnamefont {X.-D.}\ \bibnamefont
  {Wen}}, \bibinfo {author} {\bibfnamefont {L.}~\bibnamefont {Hand}}, \bibinfo
  {author} {\bibfnamefont {V.}~\bibnamefont {Labet}}, \bibinfo {author}
  {\bibfnamefont {T.}~\bibnamefont {Yang}}, \bibinfo {author} {\bibfnamefont
  {R.}~\bibnamefont {Hoffmann}}, \bibinfo {author} {\bibfnamefont {N.~W.}\
  \bibnamefont {Ashcroft}}, \bibinfo {author} {\bibfnamefont {A.~R.}\
  \bibnamefont {Oganov}}, \ and\ \bibinfo {author} {\bibfnamefont {A.~O.}\
  \bibnamefont {Lyakhov}},\ }\href {\doibase 10.1073/pnas.1103145108}
  {\bibfield  {journal} {\bibinfo  {journal} {Proc. Natl. Acad. Sci.}\ }\textbf
  {\bibinfo {volume} {108}},\ \bibinfo {pages} {6833} (\bibinfo {year}
  {2011})}\BibitemShut {NoStop}%
\bibitem [{\citenamefont {Sluiter}\ and\ \citenamefont
  {Kawazoe}(2003)}]{Sluiter2003}%
  \BibitemOpen
  \bibfield  {author} {\bibinfo {author} {\bibfnamefont {M.~H.~F.}\
  \bibnamefont {Sluiter}}\ and\ \bibinfo {author} {\bibfnamefont
  {Y.}~\bibnamefont {Kawazoe}},\ }\href {\doibase 10.1103/PhysRevB.68.085410}
  {\bibfield  {journal} {\bibinfo  {journal} {Phys. Rev. B}\ }\textbf {\bibinfo
  {volume} {68}},\ \bibinfo {pages} {085410} (\bibinfo {year}
  {2003})}\BibitemShut {NoStop}%
\bibitem [{\citenamefont {He}\ \emph {et~al.}(2012)\citenamefont {He},
  \citenamefont {Zhang}, \citenamefont {Sun}, \citenamefont {Jiao},
  \citenamefont {Zhang},\ and\ \citenamefont {Zhong}}]{He2012}%
  \BibitemOpen
  \bibfield  {author} {\bibinfo {author} {\bibfnamefont {C.}~\bibnamefont
  {He}}, \bibinfo {author} {\bibfnamefont {C.~X.}\ \bibnamefont {Zhang}},
  \bibinfo {author} {\bibfnamefont {L.~Z.}\ \bibnamefont {Sun}}, \bibinfo
  {author} {\bibfnamefont {N.}~\bibnamefont {Jiao}}, \bibinfo {author}
  {\bibfnamefont {K.~W.}\ \bibnamefont {Zhang}}, \ and\ \bibinfo {author}
  {\bibfnamefont {J.}~\bibnamefont {Zhong}},\ }\href {\doibase
  10.1002/pssr.201206358} {\bibfield  {journal} {\bibinfo  {journal} {Phys.
  Status Solidi RRL}\ }\textbf {\bibinfo {volume} {6}},\ \bibinfo {pages} {427}
  (\bibinfo {year} {2012})}\BibitemShut {NoStop}%
\bibitem [{\citenamefont {Artyukhov}\ and\ \citenamefont
  {Chernozatonskii}(2010)}]{Artyukhov2010}%
  \BibitemOpen
  \bibfield  {author} {\bibinfo {author} {\bibfnamefont {V.~I.}\ \bibnamefont
  {Artyukhov}}\ and\ \bibinfo {author} {\bibfnamefont {L.~A.}\ \bibnamefont
  {Chernozatonskii}},\ }\href {\doibase 10.1021/jp1003566} {\bibfield
  {journal} {\bibinfo  {journal} {J. Phys. Chem. A}\ }\textbf {\bibinfo
  {volume} {114}},\ \bibinfo {pages} {5389} (\bibinfo {year}
  {2010})}\BibitemShut {NoStop}%
\bibitem [{\citenamefont {Bhattacharya}\ \emph {et~al.}(2011)\citenamefont
  {Bhattacharya}, \citenamefont {Bhattacharya}, \citenamefont {Majumder},\ and\
  \citenamefont {Das}}]{Bhattacharya2011}%
  \BibitemOpen
  \bibfield  {author} {\bibinfo {author} {\bibfnamefont {A.}~\bibnamefont
  {Bhattacharya}}, \bibinfo {author} {\bibfnamefont {S.}~\bibnamefont
  {Bhattacharya}}, \bibinfo {author} {\bibfnamefont {C.}~\bibnamefont
  {Majumder}}, \ and\ \bibinfo {author} {\bibfnamefont {G.~P.}\ \bibnamefont
  {Das}},\ }\href {\doibase 10.1103/PhysRevB.83.033404} {\bibfield  {journal}
  {\bibinfo  {journal} {Phys. Rev. B}\ }\textbf {\bibinfo {volume} {83}},\
  \bibinfo {pages} {033404} (\bibinfo {year} {2011})}\BibitemShut {NoStop}%
\bibitem [{\citenamefont {Charlier}\ \emph {et~al.}(1993)\citenamefont
  {Charlier}, \citenamefont {Gonze},\ and\ \citenamefont
  {Michenaud}}]{Charlier1993}%
  \BibitemOpen
  \bibfield  {author} {\bibinfo {author} {\bibfnamefont {J.-C.}\ \bibnamefont
  {Charlier}}, \bibinfo {author} {\bibfnamefont {X.}~\bibnamefont {Gonze}}, \
  and\ \bibinfo {author} {\bibfnamefont {J.-P.}\ \bibnamefont {Michenaud}},\
  }\href {\doibase 10.1103/PhysRevB.47.16162} {\bibfield  {journal} {\bibinfo
  {journal} {Phys. Rev. B}\ }\textbf {\bibinfo {volume} {47}},\ \bibinfo
  {pages} {16162} (\bibinfo {year} {1993})}\BibitemShut {NoStop}%
\bibitem [{\citenamefont {Han}\ \emph {et~al.}(2010)\citenamefont {Han},
  \citenamefont {Yu}, \citenamefont {Merinov}, \citenamefont {van Duin},
  \citenamefont {Yazami},\ and\ \citenamefont {Goddard}}]{Han2010}%
  \BibitemOpen
  \bibfield  {author} {\bibinfo {author} {\bibfnamefont {S.~S.}\ \bibnamefont
  {Han}}, \bibinfo {author} {\bibfnamefont {T.~H.}\ \bibnamefont {Yu}},
  \bibinfo {author} {\bibfnamefont {B.~V.}\ \bibnamefont {Merinov}}, \bibinfo
  {author} {\bibfnamefont {A.~C.~T.}\ \bibnamefont {van Duin}}, \bibinfo
  {author} {\bibfnamefont {R.}~\bibnamefont {Yazami}}, \ and\ \bibinfo {author}
  {\bibfnamefont {W.~A.}\ \bibnamefont {Goddard}},\ }\href {\doibase
  10.1021/cm903760t} {\bibfield  {journal} {\bibinfo  {journal} {Chem. Mater.}\
  }\textbf {\bibinfo {volume} {22}},\ \bibinfo {pages} {2142} (\bibinfo {year}
  {2010})}\BibitemShut {NoStop}%
\bibitem [{\citenamefont {Mahajan}\ \emph {et~al.}(1974)\citenamefont
  {Mahajan}, \citenamefont {Badachhape},\ and\ \citenamefont
  {Margrave}}]{Mahajan1974}%
  \BibitemOpen
  \bibfield  {author} {\bibinfo {author} {\bibfnamefont {V.}~\bibnamefont
  {Mahajan}}, \bibinfo {author} {\bibfnamefont {R.}~\bibnamefont {Badachhape}},
  \ and\ \bibinfo {author} {\bibfnamefont {J.}~\bibnamefont {Margrave}},\
  }\href {\doibase https://doi.org/10.1016/0020-1650(74)80174-1} {\bibfield
  {journal} {\bibinfo  {journal} {Inorg. Nucl. Chem. Lett.}\ }\textbf {\bibinfo
  {volume} {10}},\ \bibinfo {pages} {1103} (\bibinfo {year}
  {1974})}\BibitemShut {NoStop}%
\bibitem [{\citenamefont {Ebert}\ \emph {et~al.}(1974)\citenamefont {Ebert},
  \citenamefont {Brauman},\ and\ \citenamefont {Huggins}}]{Ebert1974}%
  \BibitemOpen
  \bibfield  {author} {\bibinfo {author} {\bibfnamefont {L.~B.}\ \bibnamefont
  {Ebert}}, \bibinfo {author} {\bibfnamefont {J.~I.}\ \bibnamefont {Brauman}},
  \ and\ \bibinfo {author} {\bibfnamefont {R.~A.}\ \bibnamefont {Huggins}},\
  }\href {\doibase 10.1021/ja00832a054} {\bibfield  {journal} {\bibinfo
  {journal} {J. Am. Chem. Soc.}\ }\textbf {\bibinfo {volume} {96}},\ \bibinfo
  {pages} {7841} (\bibinfo {year} {1974})}\BibitemShut {NoStop}%
\bibitem [{\citenamefont {Parry}\ \emph {et~al.}(1974)\citenamefont {Parry},
  \citenamefont {Thomas}, \citenamefont {Bach},\ and\ \citenamefont
  {Evans}}]{Parry1974}%
  \BibitemOpen
  \bibfield  {author} {\bibinfo {author} {\bibfnamefont {D.}~\bibnamefont
  {Parry}}, \bibinfo {author} {\bibfnamefont {J.}~\bibnamefont {Thomas}},
  \bibinfo {author} {\bibfnamefont {B.}~\bibnamefont {Bach}}, \ and\ \bibinfo
  {author} {\bibfnamefont {E.}~\bibnamefont {Evans}},\ }\href {\doibase
  https://doi.org/10.1016/0009-2614(74)80148-X} {\bibfield  {journal} {\bibinfo
   {journal} {Chem. Phys. Lett.}\ }\textbf {\bibinfo {volume} {29}},\ \bibinfo
  {pages} {128} (\bibinfo {year} {1974})}\BibitemShut {NoStop}%
\bibitem [{\citenamefont {Touhara}\ \emph {et~al.}(1987)\citenamefont
  {Touhara}, \citenamefont {Kadono}, \citenamefont {Fujii},\ and\ \citenamefont
  {Watanabe}}]{Touhara1987}%
  \BibitemOpen
  \bibfield  {author} {\bibinfo {author} {\bibfnamefont {H.}~\bibnamefont
  {Touhara}}, \bibinfo {author} {\bibfnamefont {K.}~\bibnamefont {Kadono}},
  \bibinfo {author} {\bibfnamefont {Y.}~\bibnamefont {Fujii}}, \ and\ \bibinfo
  {author} {\bibfnamefont {N.}~\bibnamefont {Watanabe}},\ }\href {\doibase
  10.1002/zaac.19875440102} {\bibfield  {journal} {\bibinfo  {journal} {Z.
  Anorg. Allg. Chem.}\ }\textbf {\bibinfo {volume} {544}},\ \bibinfo {pages}
  {7} (\bibinfo {year} {1987})}\BibitemShut {NoStop}%
\bibitem [{\citenamefont {Sato}\ \emph {et~al.}(2004)\citenamefont {Sato},
  \citenamefont {Itoh}, \citenamefont {Hagiwara}, \citenamefont {Fukunaga},\
  and\ \citenamefont {Ito}}]{Sato2004}%
  \BibitemOpen
  \bibfield  {author} {\bibinfo {author} {\bibfnamefont {Y.}~\bibnamefont
  {Sato}}, \bibinfo {author} {\bibfnamefont {K.}~\bibnamefont {Itoh}}, \bibinfo
  {author} {\bibfnamefont {R.}~\bibnamefont {Hagiwara}}, \bibinfo {author}
  {\bibfnamefont {T.}~\bibnamefont {Fukunaga}}, \ and\ \bibinfo {author}
  {\bibfnamefont {Y.}~\bibnamefont {Ito}},\ }\href {\doibase
  https://doi.org/10.1016/j.carbon.2004.06.042} {\bibfield  {journal} {\bibinfo
   {journal} {Carbon}\ }\textbf {\bibinfo {volume} {42}},\ \bibinfo {pages}
  {2897} (\bibinfo {year} {2004})}\BibitemShut {NoStop}%
\bibitem [{\citenamefont {Cheng}\ \emph {et~al.}(2010)\citenamefont {Cheng},
  \citenamefont {Zou}, \citenamefont {Okino}, \citenamefont {Gutierrez},
  \citenamefont {Gupta}, \citenamefont {Shen}, \citenamefont {Eklund},
  \citenamefont {Sofo},\ and\ \citenamefont {Zhu}}]{Cheng2010}%
  \BibitemOpen
  \bibfield  {author} {\bibinfo {author} {\bibfnamefont {S.-H.}\ \bibnamefont
  {Cheng}}, \bibinfo {author} {\bibfnamefont {K.}~\bibnamefont {Zou}}, \bibinfo
  {author} {\bibfnamefont {F.}~\bibnamefont {Okino}}, \bibinfo {author}
  {\bibfnamefont {H.~R.}\ \bibnamefont {Gutierrez}}, \bibinfo {author}
  {\bibfnamefont {A.}~\bibnamefont {Gupta}}, \bibinfo {author} {\bibfnamefont
  {N.}~\bibnamefont {Shen}}, \bibinfo {author} {\bibfnamefont {P.~C.}\
  \bibnamefont {Eklund}}, \bibinfo {author} {\bibfnamefont {J.~O.}\
  \bibnamefont {Sofo}}, \ and\ \bibinfo {author} {\bibfnamefont
  {J.}~\bibnamefont {Zhu}},\ }\href {\doibase 10.1103/PhysRevB.81.205435}
  {\bibfield  {journal} {\bibinfo  {journal} {Phys. Rev. B}\ }\textbf {\bibinfo
  {volume} {81}},\ \bibinfo {pages} {205435} (\bibinfo {year}
  {2010})}\BibitemShut {NoStop}%
\bibitem [{\citenamefont {Mitkin}(2003)}]{Mitkin2003}%
  \BibitemOpen
  \bibfield  {author} {\bibinfo {author} {\bibfnamefont {V.~N.}\ \bibnamefont
  {Mitkin}},\ }\href {\doibase 10.1023/A:1024989132154} {\bibfield  {journal}
  {\bibinfo  {journal} {J. Struct. Chem.}\ }\textbf {\bibinfo {volume} {44}},\
  \bibinfo {pages} {82} (\bibinfo {year} {2003})}\BibitemShut {NoStop}%
\bibitem [{\citenamefont {Boukhvalov}\ \emph {et~al.}(2008)\citenamefont
  {Boukhvalov}, \citenamefont {Katsnelson},\ and\ \citenamefont
  {Lichtenstein}}]{Boukhvalov2008}%
  \BibitemOpen
  \bibfield  {author} {\bibinfo {author} {\bibfnamefont {D.~W.}\ \bibnamefont
  {Boukhvalov}}, \bibinfo {author} {\bibfnamefont {M.~I.}\ \bibnamefont
  {Katsnelson}}, \ and\ \bibinfo {author} {\bibfnamefont {A.~I.}\ \bibnamefont
  {Lichtenstein}},\ }\href {\doibase 10.1103/PhysRevB.77.035427} {\bibfield
  {journal} {\bibinfo  {journal} {Phys. Rev. B}\ }\textbf {\bibinfo {volume}
  {77}},\ \bibinfo {pages} {035427} (\bibinfo {year} {2008})}\BibitemShut
  {NoStop}%
\bibitem [{\citenamefont {\c{S}ahin}\ \emph {et~al.}(2009)\citenamefont
  {\c{S}ahin}, \citenamefont {Ataca},\ and\ \citenamefont
  {Ciraci}}]{Sahin2009}%
  \BibitemOpen
  \bibfield  {author} {\bibinfo {author} {\bibfnamefont {H.}~\bibnamefont
  {\c{S}ahin}}, \bibinfo {author} {\bibfnamefont {C.}~\bibnamefont {Ataca}}, \
  and\ \bibinfo {author} {\bibfnamefont {S.}~\bibnamefont {Ciraci}},\ }\href
  {\doibase 10.1063/1.3268792} {\bibfield  {journal} {\bibinfo  {journal}
  {Appl. Phys. Lett.}\ }\textbf {\bibinfo {volume} {95}},\ \bibinfo {pages}
  {222510} (\bibinfo {year} {2009})}\BibitemShut {NoStop}%
\bibitem [{\citenamefont {Yang}\ \emph {et~al.}(2014)\citenamefont {Yang},
  \citenamefont {Ren},\ and\ \citenamefont {Bellaiche}}]{Yang2014}%
  \BibitemOpen
  \bibfield  {author} {\bibinfo {author} {\bibfnamefont {Y.}~\bibnamefont
  {Yang}}, \bibinfo {author} {\bibfnamefont {W.}~\bibnamefont {Ren}}, \ and\
  \bibinfo {author} {\bibfnamefont {L.}~\bibnamefont {Bellaiche}},\ }\href
  {\doibase 10.1103/PhysRevB.89.245439} {\bibfield  {journal} {\bibinfo
  {journal} {Phys. Rev. B}\ }\textbf {\bibinfo {volume} {89}},\ \bibinfo
  {pages} {245439} (\bibinfo {year} {2014})}\BibitemShut {NoStop}%
\bibitem [{\citenamefont {Tang}\ and\ \citenamefont {Zhang}(2011)}]{Tang2011}%
  \BibitemOpen
  \bibfield  {author} {\bibinfo {author} {\bibfnamefont {S.}~\bibnamefont
  {Tang}}\ and\ \bibinfo {author} {\bibfnamefont {S.}~\bibnamefont {Zhang}},\
  }\href {\doibase 10.1021/jp204880f} {\bibfield  {journal} {\bibinfo
  {journal} {J. Phys. Chem. C}\ }\textbf {\bibinfo {volume} {115}},\ \bibinfo
  {pages} {16644} (\bibinfo {year} {2011})}\BibitemShut {NoStop}%
\bibitem [{\citenamefont {\c{S}ahin}\ \emph {et~al.}(2011)\citenamefont
  {\c{S}ahin}, \citenamefont {Topsakal},\ and\ \citenamefont
  {Ciraci}}]{Sahin2011}%
  \BibitemOpen
  \bibfield  {author} {\bibinfo {author} {\bibfnamefont {H.}~\bibnamefont
  {\c{S}ahin}}, \bibinfo {author} {\bibfnamefont {M.}~\bibnamefont {Topsakal}},
  \ and\ \bibinfo {author} {\bibfnamefont {S.}~\bibnamefont {Ciraci}},\ }\href
  {\doibase 10.1103/PhysRevB.83.115432} {\bibfield  {journal} {\bibinfo
  {journal} {Phys. Rev. B}\ }\textbf {\bibinfo {volume} {83}},\ \bibinfo
  {pages} {115432} (\bibinfo {year} {2011})}\BibitemShut {NoStop}%
\bibitem [{\citenamefont {Takagi}\ and\ \citenamefont
  {Kusakabe}(2002)}]{Takagi2002}%
  \BibitemOpen
  \bibfield  {author} {\bibinfo {author} {\bibfnamefont {Y.}~\bibnamefont
  {Takagi}}\ and\ \bibinfo {author} {\bibfnamefont {K.}~\bibnamefont
  {Kusakabe}},\ }\href {\doibase 10.1103/PhysRevB.65.121103} {\bibfield
  {journal} {\bibinfo  {journal} {Phys. Rev. B}\ }\textbf {\bibinfo {volume}
  {65}},\ \bibinfo {pages} {121103(R)} (\bibinfo {year} {2002})}\BibitemShut
  {NoStop}%
\bibitem [{\citenamefont {Nair}\ \emph {et~al.}(2010)\citenamefont {Nair},
  \citenamefont {Ren}, \citenamefont {Jalil}, \citenamefont {Riaz},
  \citenamefont {Kravets}, \citenamefont {Britnell}, \citenamefont {Blake},
  \citenamefont {Schedin}, \citenamefont {Mayorov}, \citenamefont {Yuan},
  \citenamefont {Katsnelson}, \citenamefont {Cheng}, \citenamefont
  {Strupinski}, \citenamefont {Bulusheva}, \citenamefont {Okotrub},
  \citenamefont {Grigorieva}, \citenamefont {Grigorenko}, \citenamefont
  {Novoselov},\ and\ \citenamefont {Geim}}]{Nair2010}%
  \BibitemOpen
  \bibfield  {author} {\bibinfo {author} {\bibfnamefont {R.~R.}\ \bibnamefont
  {Nair}}, \bibinfo {author} {\bibfnamefont {W.}~\bibnamefont {Ren}}, \bibinfo
  {author} {\bibfnamefont {R.}~\bibnamefont {Jalil}}, \bibinfo {author}
  {\bibfnamefont {I.}~\bibnamefont {Riaz}}, \bibinfo {author} {\bibfnamefont
  {V.~G.}\ \bibnamefont {Kravets}}, \bibinfo {author} {\bibfnamefont
  {L.}~\bibnamefont {Britnell}}, \bibinfo {author} {\bibfnamefont
  {P.}~\bibnamefont {Blake}}, \bibinfo {author} {\bibfnamefont
  {F.}~\bibnamefont {Schedin}}, \bibinfo {author} {\bibfnamefont {A.~S.}\
  \bibnamefont {Mayorov}}, \bibinfo {author} {\bibfnamefont {S.}~\bibnamefont
  {Yuan}}, \bibinfo {author} {\bibfnamefont {M.~I.}\ \bibnamefont
  {Katsnelson}}, \bibinfo {author} {\bibfnamefont {H.-M.}\ \bibnamefont
  {Cheng}}, \bibinfo {author} {\bibfnamefont {W.}~\bibnamefont {Strupinski}},
  \bibinfo {author} {\bibfnamefont {L.~G.}\ \bibnamefont {Bulusheva}}, \bibinfo
  {author} {\bibfnamefont {A.~V.}\ \bibnamefont {Okotrub}}, \bibinfo {author}
  {\bibfnamefont {I.~V.}\ \bibnamefont {Grigorieva}}, \bibinfo {author}
  {\bibfnamefont {A.~N.}\ \bibnamefont {Grigorenko}}, \bibinfo {author}
  {\bibfnamefont {K.~S.}\ \bibnamefont {Novoselov}}, \ and\ \bibinfo {author}
  {\bibfnamefont {A.~K.}\ \bibnamefont {Geim}},\ }\href {\doibase
  10.1002/smll.201001555} {\bibfield  {journal} {\bibinfo  {journal} {Small}\
  }\textbf {\bibinfo {volume} {6}},\ \bibinfo {pages} {2877} (\bibinfo {year}
  {2010})}\BibitemShut {NoStop}%
\bibitem [{\citenamefont {Verhoeven}\ \emph {et~al.}(2004)\citenamefont
  {Verhoeven}, \citenamefont {Dienwiebel},\ and\ \citenamefont
  {Frenken}}]{Verhoeven2004}%
  \BibitemOpen
  \bibfield  {author} {\bibinfo {author} {\bibfnamefont {G.~S.}\ \bibnamefont
  {Verhoeven}}, \bibinfo {author} {\bibfnamefont {M.}~\bibnamefont
  {Dienwiebel}}, \ and\ \bibinfo {author} {\bibfnamefont {J.~W.~M.}\
  \bibnamefont {Frenken}},\ }\href@noop {} {\bibfield  {journal} {\bibinfo
  {journal} {Phys. Rev. B}\ }\textbf {\bibinfo {volume} {70}},\ \bibinfo
  {pages} {165418} (\bibinfo {year} {2004})}\BibitemShut {NoStop}%
\bibitem [{\citenamefont {Steele}(1973)}]{Steele1972}%
  \BibitemOpen
  \bibfield  {author} {\bibinfo {author} {\bibfnamefont {W.~A.}\ \bibnamefont
  {Steele}},\ }\href {\doibase 10.1016/0039-6028(73)90264-1} {\bibfield
  {journal} {\bibinfo  {journal} {Surf. Sci.}\ }\textbf {\bibinfo {volume}
  {36}},\ \bibinfo {pages} {317} (\bibinfo {year} {1973})}\BibitemShut
  {NoStop}%
\bibitem [{Sup()}]{SupplMater}%
  \BibitemOpen
  \href@noop {} {}\bibinfo {note} {See Supplemental Material at [URL will be
  inserted by publisher] for the interlayer interaction energy of
  hydrofluorinated graphene bilayer obtained by the DFT calculations and the
  deviation of the approximation by the first Fourier harmonics from the DFT
  results as functions of relative displacements of the layers along the
  armchair and zigzag directions.}\BibitemShut {Stop}%
\bibitem [{\citenamefont {Boschetto}\ \emph {et~al.}(2013)\citenamefont
  {Boschetto}, \citenamefont {Malard}, \citenamefont {Lui}, \citenamefont
  {Mak}, \citenamefont {Li}, \citenamefont {Yan},\ and\ \citenamefont
  {Heinz}}]{Boschetto2013}%
  \BibitemOpen
  \bibfield  {author} {\bibinfo {author} {\bibfnamefont {D.}~\bibnamefont
  {Boschetto}}, \bibinfo {author} {\bibfnamefont {L.}~\bibnamefont {Malard}},
  \bibinfo {author} {\bibfnamefont {C.~H.}\ \bibnamefont {Lui}}, \bibinfo
  {author} {\bibfnamefont {K.~F.}\ \bibnamefont {Mak}}, \bibinfo {author}
  {\bibfnamefont {Z.}~\bibnamefont {Li}}, \bibinfo {author} {\bibfnamefont
  {H.}~\bibnamefont {Yan}}, \ and\ \bibinfo {author} {\bibfnamefont {T.~F.}\
  \bibnamefont {Heinz}},\ }\href {\doibase 10.1021/nl401713h} {\bibfield
  {journal} {\bibinfo  {journal} {Nano Lett.}\ }\textbf {\bibinfo {volume}
  {13}},\ \bibinfo {pages} {4620} (\bibinfo {year} {2013})}\BibitemShut
  {NoStop}%
\bibitem [{\citenamefont {Tan}\ \emph {et~al.}(2012)\citenamefont {Tan},
  \citenamefont {Han}, \citenamefont {Zhao}, \citenamefont {Wu}, \citenamefont
  {Chang}, \citenamefont {Wang}, \citenamefont {Wang}, \citenamefont {Bonini},
  \citenamefont {Marzari}, \citenamefont {Pugno}, \citenamefont {Savini},
  \citenamefont {Lombardo},\ and\ \citenamefont {Ferrari}}]{Tan2012}%
  \BibitemOpen
  \bibfield  {author} {\bibinfo {author} {\bibfnamefont {P.~H.}\ \bibnamefont
  {Tan}}, \bibinfo {author} {\bibfnamefont {W.~P.}\ \bibnamefont {Han}},
  \bibinfo {author} {\bibfnamefont {W.~J.}\ \bibnamefont {Zhao}}, \bibinfo
  {author} {\bibfnamefont {Z.~H.}\ \bibnamefont {Wu}}, \bibinfo {author}
  {\bibfnamefont {K.}~\bibnamefont {Chang}}, \bibinfo {author} {\bibfnamefont
  {H.}~\bibnamefont {Wang}}, \bibinfo {author} {\bibfnamefont {Y.~F.}\
  \bibnamefont {Wang}}, \bibinfo {author} {\bibfnamefont {N.}~\bibnamefont
  {Bonini}}, \bibinfo {author} {\bibfnamefont {N.}~\bibnamefont {Marzari}},
  \bibinfo {author} {\bibfnamefont {N.}~\bibnamefont {Pugno}}, \bibinfo
  {author} {\bibfnamefont {G.}~\bibnamefont {Savini}}, \bibinfo {author}
  {\bibfnamefont {A.}~\bibnamefont {Lombardo}}, \ and\ \bibinfo {author}
  {\bibfnamefont {A.~C.}\ \bibnamefont {Ferrari}},\ }\href {\doibase
  10.1038/nmat3245} {\bibfield  {journal} {\bibinfo  {journal} {Nat. Mater.}\
  }\textbf {\bibinfo {volume} {11}},\ \bibinfo {pages} {294} (\bibinfo {year}
  {2012})}\BibitemShut {NoStop}%
\bibitem [{\citenamefont {Lebedeva}\ \emph {et~al.}(2010)\citenamefont
  {Lebedeva}, \citenamefont {Knizhnik}, \citenamefont {Popov}, \citenamefont
  {Ershova}, \citenamefont {Lozovik},\ and\ \citenamefont
  {Potapkin}}]{Lebedeva2010}%
  \BibitemOpen
  \bibfield  {author} {\bibinfo {author} {\bibfnamefont {I.~V.}\ \bibnamefont
  {Lebedeva}}, \bibinfo {author} {\bibfnamefont {A.~A.}\ \bibnamefont
  {Knizhnik}}, \bibinfo {author} {\bibfnamefont {A.~M.}\ \bibnamefont {Popov}},
  \bibinfo {author} {\bibfnamefont {O.~V.}\ \bibnamefont {Ershova}}, \bibinfo
  {author} {\bibfnamefont {Y.~E.}\ \bibnamefont {Lozovik}}, \ and\ \bibinfo
  {author} {\bibfnamefont {B.~V.}\ \bibnamefont {Potapkin}},\ }\href {\doibase
  10.1103/PhysRevB.82.155460} {\bibfield  {journal} {\bibinfo  {journal} {Phys.
  Rev. B}\ }\textbf {\bibinfo {volume} {82}},\ \bibinfo {pages} {155460}
  (\bibinfo {year} {2010})}\BibitemShut {NoStop}%
\bibitem [{\citenamefont {Lebedeva}\ \emph
  {et~al.}(2011{\natexlab{b}})\citenamefont {Lebedeva}, \citenamefont
  {Knizhnik}, \citenamefont {Popov}, \citenamefont {Ershova}, \citenamefont
  {Lozovik},\ and\ \citenamefont {Potapkin}}]{Lebedeva2011a}%
  \BibitemOpen
  \bibfield  {author} {\bibinfo {author} {\bibfnamefont {I.~V.}\ \bibnamefont
  {Lebedeva}}, \bibinfo {author} {\bibfnamefont {A.~A.}\ \bibnamefont
  {Knizhnik}}, \bibinfo {author} {\bibfnamefont {A.~M.}\ \bibnamefont {Popov}},
  \bibinfo {author} {\bibfnamefont {O.~V.}\ \bibnamefont {Ershova}}, \bibinfo
  {author} {\bibfnamefont {Y.~E.}\ \bibnamefont {Lozovik}}, \ and\ \bibinfo
  {author} {\bibfnamefont {B.~V.}\ \bibnamefont {Potapkin}},\ }\href {\doibase
  10.1063/1.3557819} {\bibfield  {journal} {\bibinfo  {journal} {J. Chem.
  Phys.}\ }\textbf {\bibinfo {volume} {134}},\ \bibinfo {pages} {104505}
  (\bibinfo {year} {2011}{\natexlab{b}})}\BibitemShut {NoStop}%
\bibitem [{\citenamefont {Xu}\ \emph {et~al.}(2013)\citenamefont {Xu},
  \citenamefont {Li}, \citenamefont {Yakobson},\ and\ \citenamefont
  {Ding}}]{Yakobson}%
  \BibitemOpen
  \bibfield  {author} {\bibinfo {author} {\bibfnamefont {Z.}~\bibnamefont
  {Xu}}, \bibinfo {author} {\bibfnamefont {X.}~\bibnamefont {Li}}, \bibinfo
  {author} {\bibfnamefont {B.~I.}\ \bibnamefont {Yakobson}}, \ and\ \bibinfo
  {author} {\bibfnamefont {F.}~\bibnamefont {Ding}},\ }\href {\doibase
  10.1039/c3nr01854g} {\bibfield  {journal} {\bibinfo  {journal} {Nanoscale}\
  }\textbf {\bibinfo {volume} {5}},\ \bibinfo {pages} {6736} (\bibinfo {year}
  {2013})}\BibitemShut {NoStop}%
\bibitem [{\citenamefont {Sachs}\ \emph {et~al.}(2011)\citenamefont {Sachs},
  \citenamefont {Wehling}, \citenamefont {Katsnelson},\ and\ \citenamefont
  {Lichtenstein}}]{Sachs2011}%
  \BibitemOpen
  \bibfield  {author} {\bibinfo {author} {\bibfnamefont {B.}~\bibnamefont
  {Sachs}}, \bibinfo {author} {\bibfnamefont {T.~O.}\ \bibnamefont {Wehling}},
  \bibinfo {author} {\bibfnamefont {M.~I.}\ \bibnamefont {Katsnelson}}, \ and\
  \bibinfo {author} {\bibfnamefont {A.~I.}\ \bibnamefont {Lichtenstein}},\
  }\href {\doibase 10.1103/PhysRevB.84.195414} {\bibfield  {journal} {\bibinfo
  {journal} {Phys. Rev. B}\ }\textbf {\bibinfo {volume} {84}},\ \bibinfo
  {pages} {195414} (\bibinfo {year} {2011})}\BibitemShut {NoStop}%
\bibitem [{\citenamefont {Liang}\ \emph {et~al.}(2008)\citenamefont {Liang},
  \citenamefont {Sawyer}, \citenamefont {Perry}, \citenamefont {Sinnott},\ and\
  \citenamefont {Phillpot}}]{Liang2008}%
  \BibitemOpen
  \bibfield  {author} {\bibinfo {author} {\bibfnamefont {T.}~\bibnamefont
  {Liang}}, \bibinfo {author} {\bibfnamefont {W.~G.}\ \bibnamefont {Sawyer}},
  \bibinfo {author} {\bibfnamefont {S.~S.}\ \bibnamefont {Perry}}, \bibinfo
  {author} {\bibfnamefont {S.~B.}\ \bibnamefont {Sinnott}}, \ and\ \bibinfo
  {author} {\bibfnamefont {S.~R.}\ \bibnamefont {Phillpot}},\ }\href {\doibase
  10.1103/PhysRevB.77.104105} {\bibfield  {journal} {\bibinfo  {journal} {Phys.
  Rev. B}\ }\textbf {\bibinfo {volume} {77}},\ \bibinfo {pages} {104105}
  (\bibinfo {year} {2008})}\BibitemShut {NoStop}%
\bibitem [{\citenamefont {Tao}\ \emph {et~al.}(2014)\citenamefont {Tao},
  \citenamefont {Guo}, \citenamefont {Yang},\ and\ \citenamefont
  {Zhang}}]{Tao2014}%
  \BibitemOpen
  \bibfield  {author} {\bibinfo {author} {\bibfnamefont {P.}~\bibnamefont
  {Tao}}, \bibinfo {author} {\bibfnamefont {H.~H.}\ \bibnamefont {Guo}},
  \bibinfo {author} {\bibfnamefont {T.}~\bibnamefont {Yang}}, \ and\ \bibinfo
  {author} {\bibfnamefont {Z.~D.}\ \bibnamefont {Zhang}},\ }\href {\doibase
  10.1088/1674-1056/23/10/106801} {\bibfield  {journal} {\bibinfo  {journal}
  {Chin. Phys. B}\ }\textbf {\bibinfo {volume} {23}},\ \bibinfo {pages}
  {106801} (\bibinfo {year} {2014})}\BibitemShut {NoStop}%
\bibitem [{\citenamefont {Levita}\ \emph {et~al.}(2015)\citenamefont {Levita},
  \citenamefont {Molinari}, \citenamefont {Polcar},\ and\ \citenamefont
  {Righi}}]{Levita2015}%
  \BibitemOpen
  \bibfield  {author} {\bibinfo {author} {\bibfnamefont {G.}~\bibnamefont
  {Levita}}, \bibinfo {author} {\bibfnamefont {E.}~\bibnamefont {Molinari}},
  \bibinfo {author} {\bibfnamefont {T.}~\bibnamefont {Polcar}}, \ and\ \bibinfo
  {author} {\bibfnamefont {M.~C.}\ \bibnamefont {Righi}},\ }\href {\doibase
  10.1103/PhysRevB.92.085434} {\bibfield  {journal} {\bibinfo  {journal} {Phys.
  Rev. B}\ }\textbf {\bibinfo {volume} {92}},\ \bibinfo {pages} {085434}
  (\bibinfo {year} {2015})}\BibitemShut {NoStop}%
\bibitem [{\citenamefont {Lin}\ \emph {et~al.}(2013)\citenamefont {Lin},
  \citenamefont {Fang}, \citenamefont {Zhou}, \citenamefont {Lupini},
  \citenamefont {Idrobo}, \citenamefont {Kong}, \citenamefont {Pennycook},\
  and\ \citenamefont {Pantelides}}]{Lin2013}%
  \BibitemOpen
  \bibfield  {author} {\bibinfo {author} {\bibfnamefont {J.}~\bibnamefont
  {Lin}}, \bibinfo {author} {\bibfnamefont {W.}~\bibnamefont {Fang}}, \bibinfo
  {author} {\bibfnamefont {W.}~\bibnamefont {Zhou}}, \bibinfo {author}
  {\bibfnamefont {A.~R.}\ \bibnamefont {Lupini}}, \bibinfo {author}
  {\bibfnamefont {J.~C.}\ \bibnamefont {Idrobo}}, \bibinfo {author}
  {\bibfnamefont {J.}~\bibnamefont {Kong}}, \bibinfo {author} {\bibfnamefont
  {S.~J.}\ \bibnamefont {Pennycook}}, \ and\ \bibinfo {author} {\bibfnamefont
  {S.~T.}\ \bibnamefont {Pantelides}},\ }\href@noop {} {\bibfield  {journal}
  {\bibinfo  {journal} {Nano Lett.}\ }\textbf {\bibinfo {volume} {13}},\
  \bibinfo {pages} {3262} (\bibinfo {year} {2013})}\BibitemShut {NoStop}%
\bibitem [{\citenamefont {Yankowitz}\ \emph {et~al.}(2014)\citenamefont
  {Yankowitz}, \citenamefont {Wang}, \citenamefont {Birdwell}, \citenamefont
  {Chen}, \citenamefont {Watanabe}, \citenamefont {Taniguchi}, \citenamefont
  {Jacquod}, \citenamefont {San-Jose}, \citenamefont {Jarillo-Herrero},\ and\
  \citenamefont {LeRoy}}]{Yankowitz2014}%
  \BibitemOpen
  \bibfield  {author} {\bibinfo {author} {\bibfnamefont {M.}~\bibnamefont
  {Yankowitz}}, \bibinfo {author} {\bibfnamefont {J.~I.-J.}\ \bibnamefont
  {Wang}}, \bibinfo {author} {\bibfnamefont {A.~G.}\ \bibnamefont {Birdwell}},
  \bibinfo {author} {\bibfnamefont {Y.-A.}\ \bibnamefont {Chen}}, \bibinfo
  {author} {\bibfnamefont {K.}~\bibnamefont {Watanabe}}, \bibinfo {author}
  {\bibfnamefont {T.}~\bibnamefont {Taniguchi}}, \bibinfo {author}
  {\bibfnamefont {P.}~\bibnamefont {Jacquod}}, \bibinfo {author} {\bibfnamefont
  {P.}~\bibnamefont {San-Jose}}, \bibinfo {author} {\bibfnamefont
  {P.}~\bibnamefont {Jarillo-Herrero}}, \ and\ \bibinfo {author} {\bibfnamefont
  {B.~J.}\ \bibnamefont {LeRoy}},\ }\href {\doibase 10.1038/nmat3965}
  {\bibfield  {journal} {\bibinfo  {journal} {Nat. Mater.}\ }\textbf {\bibinfo
  {volume} {13}},\ \bibinfo {pages} {786} (\bibinfo {year} {2014})}\BibitemShut
  {NoStop}%
\bibitem [{\citenamefont {Lebedeva}\ \emph {et~al.}(2016)\citenamefont
  {Lebedeva}, \citenamefont {Lebedev}, \citenamefont {Popov},\ and\
  \citenamefont {Knizhnik}}]{Lebedeva2016}%
  \BibitemOpen
  \bibfield  {author} {\bibinfo {author} {\bibfnamefont {I.~V.}\ \bibnamefont
  {Lebedeva}}, \bibinfo {author} {\bibfnamefont {A.~V.}\ \bibnamefont
  {Lebedev}}, \bibinfo {author} {\bibfnamefont {A.~M.}\ \bibnamefont {Popov}},
  \ and\ \bibinfo {author} {\bibfnamefont {A.~A.}\ \bibnamefont {Knizhnik}},\
  }\href {\doibase 10.1103/PhysRevB.93.235414} {\bibfield  {journal} {\bibinfo
  {journal} {Phys. Rev. B}\ }\textbf {\bibinfo {volume} {93}},\ \bibinfo
  {pages} {235414} (\bibinfo {year} {2016})}\BibitemShut {NoStop}%
\bibitem [{\citenamefont {Lebedeva}\ and\ \citenamefont
  {Popov}(2020{\natexlab{a}})}]{Lebedeva2019a}%
  \BibitemOpen
  \bibfield  {author} {\bibinfo {author} {\bibfnamefont {I.~V.}\ \bibnamefont
  {Lebedeva}}\ and\ \bibinfo {author} {\bibfnamefont {A.~M.}\ \bibnamefont
  {Popov}},\ }\href {\doibase 10.1021/acs.jpcc.9b08306} {\bibfield  {journal}
  {\bibinfo  {journal} {J. Phys. Chem. C}\ }\textbf {\bibinfo {volume} {124}},\
  \bibinfo {pages} {2120} (\bibinfo {year} {2020}{\natexlab{a}})}\BibitemShut
  {NoStop}%
\bibitem [{\citenamefont {Lebedeva}\ and\ \citenamefont
  {Popov}(2020{\natexlab{b}})}]{Lebedeva2020}%
  \BibitemOpen
  \bibfield  {author} {\bibinfo {author} {\bibfnamefont {I.~V.}\ \bibnamefont
  {Lebedeva}}\ and\ \bibinfo {author} {\bibfnamefont {A.~M.}\ \bibnamefont
  {Popov}},\ }\href {\doibase 10.1103/PhysRevLett.124.116101} {\bibfield
  {journal} {\bibinfo  {journal} {Phys. Rev. Lett.}\ }\textbf {\bibinfo
  {volume} {124}},\ \bibinfo {pages} {116101} (\bibinfo {year}
  {2020}{\natexlab{b}})}\BibitemShut {NoStop}%
\bibitem [{\citenamefont {Dienwiebel}\ \emph {et~al.}(2004)\citenamefont
  {Dienwiebel}, \citenamefont {Verhoeven}, \citenamefont {Pradeep},
  \citenamefont {Frenken}, \citenamefont {Heimberg},\ and\ \citenamefont
  {Zandbergen}}]{Dienwiebel2004}%
  \BibitemOpen
  \bibfield  {author} {\bibinfo {author} {\bibfnamefont {M.}~\bibnamefont
  {Dienwiebel}}, \bibinfo {author} {\bibfnamefont {G.~S.}\ \bibnamefont
  {Verhoeven}}, \bibinfo {author} {\bibfnamefont {N.}~\bibnamefont {Pradeep}},
  \bibinfo {author} {\bibfnamefont {J.~W.~M.}\ \bibnamefont {Frenken}},
  \bibinfo {author} {\bibfnamefont {J.~A.}\ \bibnamefont {Heimberg}}, \ and\
  \bibinfo {author} {\bibfnamefont {H.~W.}\ \bibnamefont {Zandbergen}},\ }\href
  {\doibase 10.1103/PhysRevLett.92.126101} {\bibfield  {journal} {\bibinfo
  {journal} {Phys. Rev. Lett.}\ }\textbf {\bibinfo {volume} {92}},\ \bibinfo
  {pages} {126101} (\bibinfo {year} {2004})}\BibitemShut {NoStop}%
\bibitem [{\citenamefont {Dienwiebel}\ \emph {et~al.}(2005)\citenamefont
  {Dienwiebel}, \citenamefont {Pradeep}, \citenamefont {Verhoeven},
  \citenamefont {Zandbergen},\ and\ \citenamefont {Frenken}}]{Dienwiebel2005}%
  \BibitemOpen
  \bibfield  {author} {\bibinfo {author} {\bibfnamefont {M.}~\bibnamefont
  {Dienwiebel}}, \bibinfo {author} {\bibfnamefont {N.}~\bibnamefont {Pradeep}},
  \bibinfo {author} {\bibfnamefont {G.~S.}\ \bibnamefont {Verhoeven}}, \bibinfo
  {author} {\bibfnamefont {H.~W.}\ \bibnamefont {Zandbergen}}, \ and\ \bibinfo
  {author} {\bibfnamefont {J.~W.}\ \bibnamefont {Frenken}},\ }\href {\doibase
  10.1016/j.susc.2004.12.011} {\bibfield  {journal} {\bibinfo  {journal} {Surf.
  Sci.}\ }\textbf {\bibinfo {volume} {576}},\ \bibinfo {pages} {197} (\bibinfo
  {year} {2005})}\BibitemShut {NoStop}%
\bibitem [{\citenamefont {Filippov}\ \emph {et~al.}(2008)\citenamefont
  {Filippov}, \citenamefont {Dienwiebel}, \citenamefont {Frenken},
  \citenamefont {Klafter},\ and\ \citenamefont {Urbakh}}]{Filippov2008}%
  \BibitemOpen
  \bibfield  {author} {\bibinfo {author} {\bibfnamefont {A.~E.}\ \bibnamefont
  {Filippov}}, \bibinfo {author} {\bibfnamefont {M.}~\bibnamefont
  {Dienwiebel}}, \bibinfo {author} {\bibfnamefont {J.~W.~M.}\ \bibnamefont
  {Frenken}}, \bibinfo {author} {\bibfnamefont {J.}~\bibnamefont {Klafter}}, \
  and\ \bibinfo {author} {\bibfnamefont {M.}~\bibnamefont {Urbakh}},\ }\href
  {\doibase 10.1103/PhysRevLett.100.046102} {\bibfield  {journal} {\bibinfo
  {journal} {Phys. Rev. Lett.}\ }\textbf {\bibinfo {volume} {100}},\ \bibinfo
  {pages} {046102} (\bibinfo {year} {2008})}\BibitemShut {NoStop}%
\bibitem [{\citenamefont {van Wijk}\ \emph {et~al.}(2015)\citenamefont {van
  Wijk}, \citenamefont {Schuring}, \citenamefont {Katsnelson},\ and\
  \citenamefont {Fasolino}}]{Wijk2015}%
  \BibitemOpen
  \bibfield  {author} {\bibinfo {author} {\bibfnamefont {M.~M.}\ \bibnamefont
  {van Wijk}}, \bibinfo {author} {\bibfnamefont {A.}~\bibnamefont {Schuring}},
  \bibinfo {author} {\bibfnamefont {M.~I.}\ \bibnamefont {Katsnelson}}, \ and\
  \bibinfo {author} {\bibfnamefont {A.}~\bibnamefont {Fasolino}},\ }\href
  {\doibase 10.1088/2053-1583/2/3/034010} {\bibfield  {journal} {\bibinfo
  {journal} {2D Mater.}\ }\textbf {\bibinfo {volume} {2}},\ \bibinfo {pages}
  {034010} (\bibinfo {year} {2015})}\BibitemShut {NoStop}%
\bibitem [{\citenamefont {Gargiulo}\ and\ \citenamefont
  {Yazyev}(2018)}]{Gargiulo2018}%
  \BibitemOpen
  \bibfield  {author} {\bibinfo {author} {\bibfnamefont {F.}~\bibnamefont
  {Gargiulo}}\ and\ \bibinfo {author} {\bibfnamefont {O.~V.}\ \bibnamefont
  {Yazyev}},\ }\href {\doibase 10.1088/2053-1583/aa9640} {\bibfield  {journal}
  {\bibinfo  {journal} {2D Mater.}\ }\textbf {\bibinfo {volume} {5}},\ \bibinfo
  {pages} {015019} (\bibinfo {year} {2018})}\BibitemShut {NoStop}%
\bibitem [{\citenamefont {van Wijk}\ \emph {et~al.}(2014)\citenamefont {van
  Wijk}, \citenamefont {Schuring}, \citenamefont {Katsnelson},\ and\
  \citenamefont {Fasolino}}]{Wijk2014}%
  \BibitemOpen
  \bibfield  {author} {\bibinfo {author} {\bibfnamefont {M.~M.}\ \bibnamefont
  {van Wijk}}, \bibinfo {author} {\bibfnamefont {A.}~\bibnamefont {Schuring}},
  \bibinfo {author} {\bibfnamefont {M.~I.}\ \bibnamefont {Katsnelson}}, \ and\
  \bibinfo {author} {\bibfnamefont {A.}~\bibnamefont {Fasolino}},\ }\href
  {\doibase 10.1103/PhysRevLett.113.135504} {\bibfield  {journal} {\bibinfo
  {journal} {Phys. Rev. Lett.}\ }\textbf {\bibinfo {volume} {113}},\ \bibinfo
  {pages} {135504} (\bibinfo {year} {2014})}\BibitemShut {NoStop}%
\bibitem [{\citenamefont {Leven}\ \emph {et~al.}(2016)\citenamefont {Leven},
  \citenamefont {Maaravi}, \citenamefont {Azuri}, \citenamefont {Kronik},\ and\
  \citenamefont {Hod}}]{Leven2016}%
  \BibitemOpen
  \bibfield  {author} {\bibinfo {author} {\bibfnamefont {I.}~\bibnamefont
  {Leven}}, \bibinfo {author} {\bibfnamefont {T.}~\bibnamefont {Maaravi}},
  \bibinfo {author} {\bibfnamefont {I.}~\bibnamefont {Azuri}}, \bibinfo
  {author} {\bibfnamefont {L.}~\bibnamefont {Kronik}}, \ and\ \bibinfo {author}
  {\bibfnamefont {O.}~\bibnamefont {Hod}},\ }\href {\doibase
  10.1021/acs.jctc.6b00147} {\bibfield  {journal} {\bibinfo  {journal} {J.
  Chem. Theory Comput.}\ }\textbf {\bibinfo {volume} {12}},\ \bibinfo {pages}
  {2896} (\bibinfo {year} {2016})}\BibitemShut {NoStop}%
\bibitem [{\citenamefont {Argentero}\ \emph {et~al.}(2017)\citenamefont
  {Argentero}, \citenamefont {Mittelberger}, \citenamefont {Monazam},
  \citenamefont {Cao}, \citenamefont {Pennycook}, \citenamefont {Mangler},
  \citenamefont {Kramberger}, \citenamefont {Kotakoski}, \citenamefont {Geim},\
  and\ \citenamefont {Meyer}}]{Argentero2017}%
  \BibitemOpen
  \bibfield  {author} {\bibinfo {author} {\bibfnamefont {G.}~\bibnamefont
  {Argentero}}, \bibinfo {author} {\bibfnamefont {A.}~\bibnamefont
  {Mittelberger}}, \bibinfo {author} {\bibfnamefont {M.~R.~A.}\ \bibnamefont
  {Monazam}}, \bibinfo {author} {\bibfnamefont {Y.}~\bibnamefont {Cao}},
  \bibinfo {author} {\bibfnamefont {T.~J.}\ \bibnamefont {Pennycook}}, \bibinfo
  {author} {\bibfnamefont {C.}~\bibnamefont {Mangler}}, \bibinfo {author}
  {\bibfnamefont {C.}~\bibnamefont {Kramberger}}, \bibinfo {author}
  {\bibfnamefont {J.}~\bibnamefont {Kotakoski}}, \bibinfo {author}
  {\bibfnamefont {A.~K.}\ \bibnamefont {Geim}}, \ and\ \bibinfo {author}
  {\bibfnamefont {J.~C.}\ \bibnamefont {Meyer}},\ }\href {\doibase
  10.1021/acs.nanolett.6b04360} {\bibfield  {journal} {\bibinfo  {journal}
  {Nano Lett.}\ }\textbf {\bibinfo {volume} {17}},\ \bibinfo {pages} {1409}
  (\bibinfo {year} {2017})}\BibitemShut {NoStop}%
\bibitem [{\citenamefont {Lebedev}(2020)}]{Lebedev2020}%
  \BibitemOpen
  \bibfield  {author} {\bibinfo {author} {\bibfnamefont {A.}~\bibnamefont
  {Lebedev}},\ }\href {\doibase 10.17632/tm29mrdnbj.1} {\bibfield  {journal}
  {\bibinfo  {journal} {Mendeley Data}\ } (\bibinfo {year} {2020}),\
  10.17632/tm29mrdnbj.1},\ \bibinfo {note}
  {https://data.mendeley.com/datasets/tm29mrdnbj/1}\BibitemShut {NoStop}%
\end{thebibliography}%
\end{document}